\documentclass[longauth]{aa}

\usepackage{color}
\usepackage{graphicx}
\usepackage{natbib}
\usepackage{enumitem}
\usepackage{txfonts}
\usepackage{multirow}
\usepackage{rotating}
\usepackage{longtable}
\usepackage{supertabular,booktabs}
\usepackage{amssymb}
\usepackage{verbatim}
\usepackage{caption}
\usepackage{subcaption}
\usepackage{marvosym}
\usepackage{textcomp}
\usepackage{hyperref}
\usepackage{Euclid}
\usepackage{amsmath}

\usepackage[utf8]{inputenc}

\titlerunning{C3R2 VLT/KMOS}
\authorrunning{V. Guglielmo et al.}
\begin{document}

\title{\Euclid preparation: VIII. The Complete Calibration of the Colour--Redshift Relation survey: VLT/KMOS observations and data release}

\author{Euclid Collaboration: V.~Guglielmo$^{1}$\thanks{\email{gglvnt@mpe.mpg.de}}, R.~Saglia$^{1,2}$, F.J.~Castander$^{3,4}$, A.~Galametz$^{5}$, S.~Paltani$^{5}$, R.~Bender$^{1,2}$, M.~Bolzonella$^{6}$, P.~Capak$^{7,8}$, O.~Ilbert$^{9}$, D.C.~Masters$^{10}$
% ...
}

\institute{$^{1}$ Max Planck Institute for Extraterrestrial Physics, Giessenbachstr. 1, D-85748 Garching, Germany\\
$^{2}$ Universit\"ats-Sternwarte M\"unchen, Fakult\"at f\"ur Physik, Ludwig-Maximilians-Universit\"at M\"unchen, Scheinerstrasse 1, 81679 M\"unchen, Germany\\
$^{3}$ Institute of Space Sciences (ICE, CSIC), Campus UAB, Carrer de Can Magrans, s/n, 08193 Barcelona, Spain\\
$^{4}$ Institut d’Estudis Espacials de Catalunya (IEEC), 08034 Barcelona, Spain\\
$^{5}$ Department of Astronomy, University of Geneva, ch. d'\'Ecogia 16, CH-1290 Versoix, Switzerland\\
$^{6}$ INAF-Osservatorio di Astrofisica e Scienza dello Spazio di Bologna, Via Piero Gobetti 93/3, I-40129 Bologna, Italy\\
$^{7}$ California institute of Technology, 1200 E California Blvd, Pasadena, CA 91125, USA\\
$^{8}$ Cosmic Dawn Center (DAWN), Niels Bohr Institute, University of Copenhagen, Vibenshuset, Lyngbyvej 2, DK-2100 Copenhagen, Denmark\\
$^{9}$ Aix-Marseille Univ, CNRS, CNES, LAM, Marseille, France\\
$^{10}$ Infrared Processing and Analysis Center, California Institute of Technology, Pasadena, CA 91125, USA\\
$^{11}$ Jet Propulsion Laboratory, California Institute of Technology, 4800 Oak Grove Drive, Pasadena, CA, 91109, USA\\
$^{12}$ INAF-Osservatorio Astronomico di Brera, Via Brera 28, I-20122 Milano, Italy\\
$^{13}$ Universidad de la Laguna, E-38206, San Crist\'{o}bal de La Laguna, Tenerife, Spain\\
$^{14}$ Instituto de Astrof\'{i}sica de Canarias. Calle V\'{i}a L\`{a}ctea s/n, 38204, San Crist\'{o}bal de la Laguna, Tenerife, Spain\\
$^{15}$ Dipartimento di Fisica e Astronomia, Universit\'a di Bologna, Via Gobetti 93/2, I-40129 Bologna, Italy\\
$^{16}$ INFN-Sezione di Bologna, Viale Berti Pichat 6/2, I-40127 Bologna, Italy\\
$^{17}$ IFPU, Institute for Fundamental Physics of the Universe, via Beirut 2, 34151 Trieste, Italy\\
$^{18}$ INAF-Osservatorio Astronomico di Trieste, Via G. B. Tiepolo 11, I-34131 Trieste, Italy\\
$^{19}$ INAF-Osservatorio Astrofisico di Torino, Via Osservatorio 20, I-10025 Pino Torinese (TO), Italy\\
$^{20}$ Department of Mathematics and Physics, Roma Tre University, Via della Vasca Navale 84, I-00146 Rome, Italy\\
$^{21}$ Institut de Recherche en Astrophysique et Plan\'etologie (IRAP), Universit\'e de Toulouse, CNRS, UPS, CNES, 14 Av. Edouard Belin, F-31400 Toulouse, France\\
$^{22}$ INAF-Osservatorio Astronomico di Capodimonte, Via Moiariello 16, I-80131 Napoli, Italy\\
$^{23}$ INFN-Bologna, Via Irnerio 46, I-40126 Bologna, Italy\\
$^{24}$ Dipartimento di Fisica e Scienze della Terra, Universit\'a degli Studi di Ferrara, Via Giuseppe Saragat 1, I-44122 Ferrara, Italy\\
$^{25}$ INAF, Istituto di Radioastronomia, Via Piero Gobetti 101, I-40129 Bologna, Italy\\
$^{26}$ INFN-Sezione di Torino, Via P. Giuria 1, I-10125 Torino, Italy\\
$^{27}$ Dipartimento di Fisica, Universit\'a degli Studi di Torino, Via P. Giuria 1, I-10125 Torino, Italy\\
$^{28}$ Universit\'e C\^ote d'Azur, Observatoire de la C\^ote d’Azur, CNRS, Laboratoire Lagrange, France\\
$^{29}$ INAF-IASF Milano, Via Alfonso Corti 12, I-20133 Milano, Italy\\
$^{30}$ Institut de F\'isica d’Altes Energies IFAE, 08193 Bellaterra, Barcelona, Spain\\
$^{31}$ Instituto de Astrof\'isica e Ci\^encias do Espa\c{c}o, Faculdade de Ci\^encias, Universidade de Lisboa, Tapada da Ajuda, PT-1349-018 Lisboa, Portugal\\
$^{32}$ AIM, CEA, CNRS, Universit\'{e} Paris-Saclay, Universit\'{e} Paris Diderot, Sorbonne Paris Cit\'{e}, F-91191 Gif-sur-Yvette, France\\
$^{33}$ INAF-Osservatorio Astronomico di Roma, Via Frascati 33, I-00078 Monteporzio Catone, Italy\\
$^{34}$ Observatoire de Sauverny, Ecole Polytechnique F\'ed\'erale de Lau- sanne, CH-1290 Versoix, Switzerland\\
$^{35}$ Department of Physics "E. Pancini", University Federico II, Via Cinthia 6, I-80126, Napoli, Italy\\
$^{36}$ INFN section of Naples, Via Cinthia 6, I-80126, Napoli, Italy\\
$^{37}$ INAF-Osservatorio Astrofisico di Arcetri, Largo E. Fermi 5, I-50125, Firenze, Italy\\
$^{38}$ Centre National d'Etudes Spatiales, Toulouse, France\\
$^{39}$ Institute for Astronomy, University of Edinburgh, Royal Observatory, Blackford Hill, Edinburgh EH9 3HJ, UK\\
$^{40}$ University of Nottingham, University Park, Nottingham NG7 2RD, UK\\
$^{41}$ European Space Agency/ESRIN, Largo Galileo Galilei 1, 00044 Frascati, Roma, Italy\\
$^{42}$ ESAC/ESA, Camino Bajo del Castillo, s/n., Urb. Villafranca del Castillo, 28692 Villanueva de la Ca\~nada, Madrid, Spain\\
$^{43}$ Univ Lyon, Univ Claude Bernard Lyon 1, CNRS/IN2P3, IP2I Lyon, UMR 5822, F-69622, Villeurbanne, France\\
$^{44}$ University of Lyon, UCB Lyon 1, CNRS/IN2P3, IUF, IP2I Lyon, France\\
$^{45}$ Mullard Space Science Laboratory, University College London, Holmbury St Mary, Dorking, Surrey RH5 6NT, UK\\
$^{46}$ Departamento de F\'isica, Faculdade de Ci\^encias, Universidade de Lisboa, Edif\'icio C8, Campo Grande, PT1749-016 Lisboa, Portugal\\
$^{47}$ Instituto de Astrof\'isica e Ci\^encias do Espa\c{c}o, Faculdade de Ci\^encias, Universidade de Lisboa, Campo Grande, PT-1749-016 Lisboa, Portugal\\
$^{48}$ Department of Physics, Oxford University, Keble Road, Oxford OX1 3RH, UK\\
$^{49}$ INFN-Padova, Via Marzolo 8, I-35131 Padova, Italy\\
$^{50}$ School of Physics, HH Wills Physics Laboratory, University of Bristol, Tyndall Avenue, Bristol, BS8 1TL, UK\\
$^{51}$ Department of Physics, P.O. Box 64, 00014 University of Helsinki, Finland\\
$^{52}$ Department of Physics and Helsinki Institute of Physics, Gustaf H\"allstr\"omin katu 2, 00014 University of Helsinki, Finland\\
$^{53}$ Dipartimento di Fisica "Aldo Pontremoli", Universit\'a degli Studi di Milano, Via Celoria 16, I-20133 Milano, Italy\\
$^{54}$ INFN-Sezione di Milano, Via Celoria 16, I-20133 Milano, Italy\\
$^{55}$ Astronomisches Institut, Ruhr-Universit\"at Bochum, Universit\"atsstr. 150, 44801 Bochum, Germany\\
$^{56}$ Leiden Observatory, Leiden University, Niels Bohrweg 2, 2333 CA Leiden, The Netherlands\\
$^{57}$ von Hoerner \& Sulger GmbH, Schlo{\ss}Platz 8, D-68723 Schwetzingen, Germany\\
$^{58}$ Max-Planck-Institut f\"ur Astronomie, K\"onigstuhl 17, D-69117 Heidelberg, Germany\\
$^{59}$ Aix-Marseille Univ, CNRS/IN2P3, CPPM, Marseille, France\\
$^{60}$ Institut d'Astrophysique de Paris, 98bis Boulevard Arago, F-75014, Paris, France\\
$^{61}$ Institut de Physique Nucl\'eaire de Lyon, 4, rue Enrico Fermi, 69622, Villeurbanne cedex, France\\
$^{62}$ Universit\'e de Gen\`eve, D\'epartement de Physique Th\'eorique and Centre for Astroparticle Physics, 24 quai Ernest-Ansermet, CH-1211 Gen\`eve 4, Switzerland\\
$^{63}$ European Space Agency/ESTEC, Keplerlaan 1, 2201 AZ Noordwijk, The Netherlands\\
$^{64}$ Institute of Theoretical Astrophysics, University of Oslo, P.O. Box 1029 Blindern, N-0315 Oslo, Norway\\
$^{65}$ NOVA optical infrared instrumentation group at ASTRON, Oude Hoogeveensedijk 4, 7991PD, Dwingeloo, The Netherlands\\
$^{66}$ Institute of Cosmology and Gravitation, University of Portsmouth, Portsmouth PO1 3FX, UK\\
$^{67}$ Argelander-Institut f\"ur Astronomie, Universit\"at Bonn, Auf dem H\"ugel 71, 53121 Bonn, Germany\\
$^{68}$ Centre for Extragalactic Astronomy, Department of Physics, Durham University, South Road, Durham, DH1 3LE, UK\\
$^{69}$ Universit\'{e} C\^{o}te d'Azur, Observatoire de la C\^{o}te d'Azur, CNRS, Laboratoire Lagrange, Bd de l'Observatoire, CS 34229, 06304 Nice cedex 4, France\\
$^{70}$ Universit\'{e} de Paris, F-75013, Paris, France, LERMA, Observatoire de Paris, PSL Research University, CNRS, Sorbonne Universit\'e, F-75014 Paris, France\\
$^{71}$ Istituto Nazionale di Astrofisica (INAF) - Osservatorio di Astrofisica e Scienza dello Spazio (OAS), Via Gobetti 93/3, I-40127 Bologna, Italy\\
$^{72}$ INAF-IASF Bologna, Via Piero Gobetti 101, I-40129 Bologna, Italy\\
$^{73}$ INFN, Sezione di Lecce, Via per Arnesano, CP-193, I-73100, Lecce, Italy\\
$^{74}$ Department of Mathematics and Physics E. De Giorgi, University of Salento, Via per Arnesano, CP-I93, I-73100, Lecce, Italy\\
$^{75}$ Dipartimento di Fisica e Astronomia “G.Galilei", Universit\'a di Padova, Via Marzolo 8, I-35131 Padova, Italy\\
$^{76}$ Departamento de F\'isica, FCFM, Universidad de Chile, Blanco Encalada 2008, Santiago, Chile\\
$^{77}$ Centro de Investigaciones Energ\'eticas, Medioambientales y Tecnol\'ogicas (CIEMAT), Avenida Complutense 40, 28040 Madrid, Spain\\
$^{78}$ Universidad Polit\'ecnica de Cartagena, Departamento de Electr\'onica y Tecnolog\'ia de Computadoras, 30202 Cartagena, Spain\\
$^{79}$ Jodrell Bank Centre for Astrophysics, School of Physics and Astronomy, University of Manchester, Oxford Road, Manchester M13 9PL, UK\\
$^{80}$ Department of Physics and Astronomy, University College London, Gower Street, London WC1E 6BT, UK\\
}

\abstract{
The Complete Calibration of the Colour--Redshift Relation survey (C3R2) is a spectroscopic effort involving ESO and Keck facilities designed specifically to empirically calibrate the galaxy colour--redshift relation --- $P(z\textbar \vec{C})$ to the {\it Euclid} depth ($i_{\rm AB}=24.5$) and is intimately linked to the success of upcoming  Stage IV dark energy missions based on weak lensing cosmology. 
The aim is to build a spectroscopic calibration sample that is as representative as possible of the galaxies of the {\it Euclid} weak lensing sample.
In order to minimise the number of spectroscopic observations necessary to fill the gaps in current knowledge of the $P(z\textbar \vec{C})$, self-organising map (SOM) representations of the galaxy colour space have been constructed.
Here we present the first results of an ESO\MVAt VLT Large Programme approved in the context of C3R2, which makes use of the two VLT optical and near-infrared multi-object spectrographs, FORS2 and KMOS.
This data release paper focuses on high-quality spectroscopic redshifts of high-redshift galaxies observed with the KMOS spectrograph in the near-infrared $H$- and $K$-bands.
A total of 424 highly-reliable redshifts are measured in the $1.3\leq z\leq 2.5 $ range, with total success rates of 60.7\% in the $H$-band and 32.8\% in the $K$-band. The newly determined redshifts fill 55\% of high (mainly regions with no spectroscopic measurements) and 35\% of lower (regions with low-resolution/low-quality spectroscopic measurements) priority empty SOM grid cells.
We measured H$\alpha$ fluxes in a $1\farcs 2$ radius aperture from the spectra of the spectroscopically confirmed galaxies and converted them into star formation rates.
In addition, we performed an SED fitting analysis on the same sample in order to derive stellar masses, $E(B-V)$, total magnitudes, and SFRs. We combine the results obtained from the spectra with those derived via SED fitting, and we show that the spectroscopic failures come from either weakly star-forming galaxies (at $z<1.7$, i.e. in the $H$-band) or low S/N spectra (in the $K$-band) of $z>2$ galaxies.}



\date{Received xxx; accepted yyy}

\maketitle

\section{Introduction}



The existence of a direct connection between cosmic shear and the presence of gravitational fields created by the distribution of matter along the line of sight motivated the development of a number of weak lensing cosmological surveys. These are both space based, such as {\it Euclid} \citep{Laureijs2011} and WFIRST \citep{Spergel2015}, and ground based, such as the ongoing Kilo-Degree Survey \citep[KiDS,][]{deJong2013}, Dark Energy Survey \citep[DES,][]{DES2016}, Hyper Suprime-Cam Subaru Strategic Programme \citep[HSC SSP,][]{Aihara2017}, and the future Vera C. Rubin Observatory survey \citep[LSST,][]{LSST2009}.
The main advantage of space missions with respect to ground-based ones is the absence of atmospheric turbulence, which leads to images with smaller and more stable point-spread functions (PSFs), allowing cosmological analyses at higher redshifts.
Besides turbulence, space is key for near-infrared observations, thanks to the lower background, which makes it possible to reach higher redshift than the ground-based surveys.

The aims of the aforementioned projects are to determine galaxy shape distortions, make use of weak lensing principles to measure the geometry of the Universe, and trace the evolution of large-scale structure (LSS) to shed light on the complex relation between galaxies and the dark components of the Universe.
In this respect, the outcome of these ambitious programmes heavily depends on the precise determination of the true ensemble redshift distribution, or $N(z)$, and thus an accurate reconstruction of the 3D distribution of galaxies. To the lowest order, weak lensing is primarily sensitive to the mean redshift and the width of the redshift distribution in tomographic bins \citep{Amara2007}.

Moreover, the sensitivity of weak lensing tomography to the dark energy equation of state cannot disregard the ability to measure the growth of structure by dividing the source samples by redshift.
The difficulty of finding optimal tomographic redshift bins for cosmic shear analysis has been studied in recent works, and solutions based on dimensionality reduction approach through self-organising maps \citep[SOM,][]{Kohonen2001} have been explored \citep{Kitching2019}.

In the case of {\it Euclid}, this translates into stringent requirements on the knowledge of the redshift distribution of sources evaluated in terms of (1) the precision of individual redshifts, which must be $\sigma_z < 0.05(1+z)$, and (2) the mean redshift $\langle z\rangle$ of each tomographic bin, which must be constrained at the level of $\Delta \langle z\rangle \leq 0.002(1+\langle z\rangle)$.

The {\it Euclid} satellite, scheduled for launch in 2022, will observe galaxies out to at least $z=2$ over $15\,000\,\rm deg^2$ by means of two instruments: VIS, an optical imager that will reach an AB magnitude depth of 24.5 with a single broad $r+i+z$ filter, and NISP, a combined near-infrared imager (in $Y$, $J$ and $H$) and slitless spectrograph.
The estimated number of weak lensing source galaxies that will be imaged from {\it Euclid} makes their systematic spectroscopic follow-up unfeasible; this mission is thus critically dependent upon the determination of accurate photometric redshifts ($z_{\rm phot}$). However, the accuracy of current photometric redshifts based on multi-band optical surveys is to the order of $\sigma_z/(1+z) = 0.03-0.06,$ and the fraction of catastrophic outliers --- defined as objects whose $z_{\rm phot}$ differs from their spectroscopic redshift ($z_{\rm spec}$) by more than $0.15(1+z)$ is to the order of a few tens of percent \citep{Ma2006,Hildebrandt2010}.
While small changes in $z_{\rm phot}$ precision per source have a relatively small impact on cosmological parameter estimates, small systematic errors in $z_{\rm phot}$ can dominate all other uncertainties for these experiments.

In this work, we present the results of all the redshift measurements on $z>1$ galaxies performed during five semesters in the context of an ESO Large Programme at the Very Large Telescope (VLT, the detailed presentation can be found in Sect. \ref{Sec:c3r2}), using the near infrared KMOS spectrograph. The campaign conducted with FORS2 on the lower redshift targets will be presented in a companion paper (Castander et al., in prep.).
The paper is organised as follows: Sect. \ref{Sec:c3r2} presents the concept and the characteristics of the C3R2 survey; in Sect. \ref{Sec:target_selection}, we present the survey strategy; in Sect. \ref{Sec:observations_data_reduction} we describe the observations and data reduction; in Sect. \ref{Sec:redshift_assignment}, we discuss the redshift determination and the attribution of a flagging scheme consistent over the whole C3R2 survey; in Sect. \ref{Sec:results_I}, we present the results of the redshift assignment in terms of success rate and SOM cell coverage; in Sect. \ref{Sec:results_II}, we determine and discuss the galaxy physical properties in terms of H$_\alpha$ fluxes and stellar masses and investigate their location in the star formation rate stellar mass (SFR-$M_\star$) plane; finally, we present our conclusions in Sect. \ref{Sec:conclusions}.

Throughout the paper, we assume $H_0 = 70 \, \si{\kmsMpc}$, $\Omega_{\rm m}=0.3$, $\Omega_{\Lambda}=0.7$. We adopt a \cite{Chabrier2003} initial mass function (IMF) in the mass range 0.1 -- 100\,$\si{\solarmass}$.

\section{Mapping the colour--redshift relation with spectroscopy}
\label{Sec:c3r2}
In order to overcome the limitations of current techniques used to estimate n(z), a complete calibration set of spectroscopic data is required. This spectroscopic calibration sample should be representative of the entire range of galaxy types and redshifts that are going to be exploited by a given weak lensing survey.

\subsection{Dimensionality reduction approach to $P(z\textbar \vec{C})$ calibration}

In order to shed light on our current knowledge of the galaxy population for weak lensing measurements, and in particular for {\it Euclid}, \citet[hereafter M15]{Masters2015} made use of a SOM to map the high-dimensional galaxy colour space onto a 2D plane.
We used the SOM to group galaxies according to the similarity of their colours (i.e. of their spectral energy distributions; SEDs) in order to unveil which regions of the galaxy colour space (represented by cells in the plane) are not represented in currently available spectroscopic surveys. This grouping strategy allows us, in turn, to minimise the number of additional spectroscopic redshifts necessary to build a complete and representative calibration sample. The underlying assumption of this methodology is that, for a dense enough SOM and a sufficiently high-dimensional colour space, there is a unique and non-degenerate relation between the position occupied by a galaxy in a multi-colour space and its redshift --- $P(z\textbar \vec{C})$.
Similar dimensionality reduction approaches in the context of weak lensing cosmological surveys have been used in recent works, using, for example, absolute magnitudes instead of colours in order to calibrate photometric redshifts \citep[][in press]{Wright2019}. The authors stress the importance of using magnitudes as a reference to an absolute flux scale in order to calibrate the n(z) for {\it Euclid}.
Starting from a photometric sample of galaxies selected using the {\it Euclid} magnitude limit and grouped using the {\it Euclid} colours and the corresponding spectroscopic sub-samples available in the Cosmological Evolution Survey (COSMOS, \citealt{Scoville2007}) field, M15 estimated that a total collection of $\sim 10-15\,{\rm K}$ spectra would be necessary in order to fill the galaxy colour space and cover the whole set of parameters characterising the galaxy population that will be observed by {\it Euclid}. About half of them are already available from various spectroscopic surveys in the literature, whereas approximately 5000 new redshifts should be observed in order to calibrate the current photometric redshift techniques and meet the {\it Euclid} requirements.
Galaxies in these unexplored regions of colour space are generally fainter than $i_{\rm AB}\sim23$ and lie at intermediate redshift, $0.2 <z<2.0$; they correspond to a population of faint, blue galaxies at intermediate redshift, which have not been targeted because they are near the magnitude limit of previous surveys.  However, their abundance and unique colours make them an important part of the galaxy population and crucial sources for weak lensing cosmology. Based on their spectral energy distributions, we expect the objects targeted to be mostly low-metallicity galaxies with strong emission lines. A minor number of cells contain faint red galaxies that are either passively evolving or dust obscured, but these constitute only 10--20\% of the unexplored sample.
Hence, M15 collected a large number of existing spectroscopic measurements in the COSMOS field \citep{Capak2007,Scoville2007,Lilly2007} to identify the type (and number) of sources that require spectroscopic follow-up in order to accurately map the full colour-redshift relation of galaxies. The work has since then been extended to four additional fields: the VIMOS VLT Deep Survey (VVDS) field, the Subaru/\XMMN Deep Survey (SXDF) field, the Extended Groth Strip field (EGS, within the All-Wavelength EGS International Survey, AEGIS), and the Extended {\it Chandra} Deep Field-South (E-CDFS) field.

\subsection{C3R2 overview}

The Complete Calibration of the Color--Redshift Relation (C3R2; \citealt{Masters2017}; M17 hereafter) survey was designed to perform a systematic spectroscopic effort by means of two observing campaigns involving two telescope facilities. Part of the spectroscopic follow-up is conducted with the Keck telescopes using a combination of the DEIMOS, LRIS, and MOSFIRE instruments, with time allocated from all Keck partners (M17). The second part is overseen by the ESO Very Large Telescope (VLT) and its UT1 instruments FORS2 and KMOS.

M17 presented the results of the first five nights of observations using the Keck facilities during the 2016A semester, leading to the release of 1283 high-confidence redshifts (Data Release 1). A further 3171 new high-quality spectroscopic redshifts were obtained during 2016B and 2017A semesters and are released in \cite[][M19, Data Release 2]{Masters2019}. A third C3R2\MVAt Keck data release is in preparation (Stanford et al., in prep.).

\subsection{C3R2 \MVAt VLT}

In order to build a large sample of spectroscopic redshifts for the calibration of the photometric redshifts of upcoming cosmological surveys we obtained a $\rm 200\,h$ large programme (199.A-0732; PI F. J. Castander) in service mode over four semesters (Period P99: 1$^{\rm st}$ April 2017 -- P102: 31$^{\rm st}$ March 2019 $+ \, carryover$). The large programme allocated $\rm 112\,h$ to FORS2, a multi-object optical slit spectrograph and $\rm 88.8\,h$ to KMOS, an integral field unit (IFU) spectrograph covering the near-infrared wavelength regime.
KMOS observations were automatically carried over P103 to complete a few P102 pointings in the SXDF field.
The VLT campaign targets the same extragalactic fields observed with the Keck programme with the exception of EGS, which is not accessible from the southern hemisphere.

\section{Target selection and KMOS IFU settings}
\label{Sec:target_selection}

\subsection{Observed fields}

In order to reduce the impact of sample variance on the calibration of photometric redshifts, the spectroscopic follow-up observations are conducted in a number of extragalactic calibrations and deep fields planned for the {\it Euclid} mission. However, we expect these commonly observed fields to also be the calibration fields of other upcoming surveys such as LSST and WFIRST; this spectroscopic follow-up effort will therefore be beneficial for the wide field survey community at large.

The major driving criterion in the choice of such fields is the possibility of collecting a homogeneous and well-calibrated photometric sample of galaxies observed in eight filters ($ugrizYJH$, seven colours) from the optical to the near-infrared domain down to the {\it Euclid} limiting magnitude but with five times higher signal-to-noise ratio.
A combination of the Canada-France-Hawaii Telescope Legacy Survey (CFHTLS) deep fields in the $ugriz$ optical magnitude and the VISTA or CFHT-WIRCAM Deep Survey (WIRDS) in the $YHK$ near-infrared bands was found to meet these requirements. 
The finally targeted fields are COSMOS (from which the SOM was derived; RA=\ra{10;00;}, Dec=\ang{02;12;}), the VIMOS-VLT Deep Survey field centred at RA=\ra{02;;} (VVDS-02h, VVDS hereafter; \citealt{LeFevre2005}; RA=\ra{02;26;} Dec=\ang{-04;30;}), the Subaru/\XMMN Deep Survey field (SXDF; \citealt{Furusawa2008}; RA=\ra{02;18;} Dec=\ang{-05;;}), and the Extended $Chandra$ Deep Field-South Survey field (ECDFS; \citealt{Lehmer2005}; four fields centred at the following coordinates: Field 1, RA=\ra{03;33;05.61} Dec=\ang{-27;41;08.84}; Field 2, RA=\ra{03;31;51.43} Dec=\ang{-27;41;38.80}; Field 3, RA=\ra{03;31;49.94} Dec=\ang{-27;57;14.56}; Field 4, RA=\ra{03;33;02.93} Dec=\ang{-27;57;16.08}). The Keck part of C3R2 additionally targets the Extended Groth Strip field (EGS; RA=\ra{14;19;} Dec=\ang{52;41;}), inaccessible to VLT facilities.
We note that the SXDF and E-CDFS fields currently lack uniform photometry in the full suite of the aforementioned optical and near-infrared filters at the required depth, but as they provide a considerable number of spectroscopic redshifts, they were included after applying a rough colour correction to convert into the CFHTLS+VISTA/WIRDS-like system (see M17).

\subsection{Prioritisation scheme and target selection}
\label{Sec:prioritization_scheme}

C3R2 prioritises targets in regions of the SOM that lack spectroscopic redshifts. High-priority targets have colours that are frequent (i.e. fall in cells with high occupation) and are therefore extremely valuable in calibrating the redshift-to-colour relation.
The C3R2 prioritisation scheme (extensively described in M19) therefore gives higher weights to sources with common colours in still uncharted cells. As observations are obtained and spectroscopic redshifts determined, the target catalogue and priority flags are updated.

\begin{figure}
    \centering
    \includegraphics[scale=0.18]{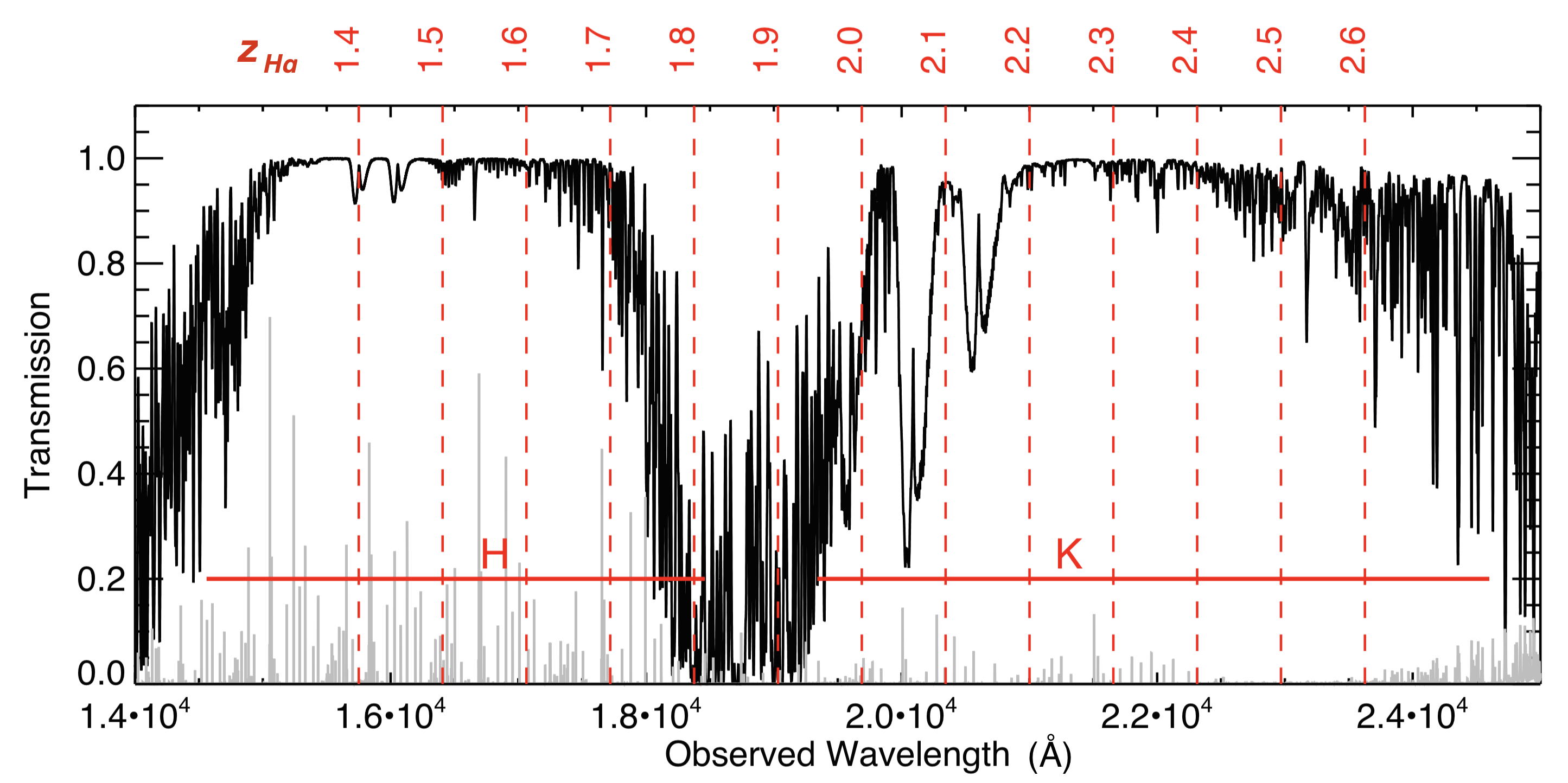}
    \caption{Telluric absorption curve (black curve) in wavelength range covered by the KMOS $H$- and $K$-band gratings (red horizontal lines); the light grey spectrum in the bottom part of the panel represents the emission lines produced by the OH radical in the atmosphere between 0.61\,\micron \, and 2.62\,\micron. The red labels on the top horizontal axis indicate the redshift ($1.4<z<2.6$) of a galaxy whose H$\alpha$ emission line falls at the wavelength indicated by the position of the vertical red dashed lines.}
    \label{fig:transmission_curve_IR}
\end{figure}

Spectroscopic redshift measurements are based on the identification of emission lines in the observed galaxy spectra, with higher priority given to the detection of the often prominent H$\alpha$ line ($\lambda \, 6564.61$\,\AA \footnote{In order to operate at near-infrared wavelengths, the entire working parts of the instrument are cooled to below $-130$\textdegree C with the detector cooled even further to below $-200$\textdegree C. To achieve this, the entire instrument is contained in a vacuum within a cryostat to prevent icing and extra heat load on the fragile components. Therefore, the wavelength values should be considered in rest-frame vacuum units (e.g. ${\rm H}\alpha_{\rm restframe}$=6564.61\,\AA).}). 
The grisms selected for the KMOS observations are $H$ (1.456 -- 1.846\,\micron) and $K$ (1.934 -- 2.460\,\micron); we thus target galaxies with a photometric redshift that positions the H$\alpha$ line within the observed wavelength range but avoids its contamination by atmospheric absorption windows as well as OH night-sky emission lines, as shown in Fig. \ref{fig:transmission_curve_IR}.

We selected high-redshift star-forming galaxy candidates with $1.3 < z_{\rm phot} < 1.7$ and $2.0 < z_{\rm phot} < 2.5$ to be observed with the $H$ and $K$ grisms, respectively, and divide them into two classes based on the prioritisation scheme defined in M19:

\begin{itemize}

\item[]- $H$-band, priority 1: $1.3 \leq z_{\rm phot} \rm \leq 1.7$, $i_{\rm tot} \leq 24.5$, and the priority flag computed in M19 ($P_{\rm F}$) $\geq 500$\footnote{The $P_{\rm F}$ parameter computed in M19 ranges from 0 up to 3750; 89\% of the SOM cells have $P_{\rm F}\leq 500$.};

\item[]- $H$-band, priority 2: $1.3 \leq z_{\rm phot} \rm \leq 1.7$, $i_{\rm tot} \leq 24.5$, and $200 \leq P_{\rm F}< 500$;

\item[]- $K$-band, priority 1: $2.0 \leq z_{\rm phot} \leq 2.5$, $i_{\rm tot} \leq 24.7$, and $P_{\rm F} \geq 500$;

\item[]- $K$-band, priority 2: $2.0 \leq z_{\rm phot} \leq 2.5$, $i_{\rm tot} \leq 24.7$, and $200 \leq P_{\rm F} < 500$.

\end{itemize}

The $H$-band $P_{\rm F} \geq 500$ corresponds to the top $7.2\%$ of KMOS selection list, $P_{\rm F} \geq 200$ corresponds to the top $18\%$.
$K$-band priority $>500$ corresponds to the top $16\%$ of the KMOS selection list, priority $>200$ corresponds to the top $33\%$.

A fraction of the COSMOS, SXDF, and E-CDFS fields have been extensively observed in the past with KMOS as part of the KMOS3D programme, one of the KMOS Guarantee Time Observations programmes \citep{Wilkinson2015} using the $YJ$, $H,$ and $K$ gratings. We removed all sources already observed by the KMOS3D team from the present target selection. Their spectroscopic redshifts (of exquisite precision) are available publicly \citep{Wisnioski2019} and are going to be used for the calibration of the {\it Euclid} photometric redshifts (KMOS3D, \url{http://www.mpe.mpg.de/ir/KMOS3D/data}).

\section{Observations and data reduction}
\label{Sec:observations_data_reduction}
In this section, we describe the acquisition and reduction of the data.

\subsection{Observation design}
KMOS is a multiplexed near-infrared integral field system (IFS) with $24$ deployable image slicers (commonly referred to as `arms'), surveying a $7^\prime.2$ diameter patrol field area. Each arm has a field of view (FoV) of 2\farcs8 $\times$ 2\farcs8 (14 $\times$ 14 pixel IFS units) and a spatial resolution of 0\farcs2/spaxel. The IFS units connect to three cryogenic grating spectrometers with 2k$\times$2k Hawaii-2RG HgCdTe detectors. As previously mentioned, among the five available KMOS gratings ($IZ$, $YJ$, $H$, $K$, $HK$), our observations make use of the $H$- and $K$-bands (plus tentative $YJ$), characterised by a typical spectral resolution of about 3500. The observations were prepared with the KMOS ARM Allocator (KARMA; \citealt{Wegner2008}) software, and submitted through the Phase 2 Proposal Preparation (P2PP) tool. Hereafter an individual KARMA setup (made of 24 arm allocations) is referred to as a `pointing'. 
Each pointing was observed for a total of 3600\,s split into single exposures of 300\,s each, using an O-S-O-O-S-O pattern (i.e. a `sky' exposure is observed every two `object' exposures). The sky exposures were offset with respect to targets to the closest position uncontaminated by sources. Additional sub-pixel/pixel dithering shifts were also applied at every exposure to minimise the impact of pixel-to-pixel variation and bad pixels in the final science data cube.
One of the $24$ KMOS IFUs was allocated to a star (with an observed magnitude of $15.0<H<16.5$) 
during the science observations (with the exception of 7/36 pointings). The star allows us to track variations in the PSF and photometric conditions between the frames; the star is therefore referred to as the PSF star. 

The standard requirements of the KARMA software for preparing a KMOS pointing are, firstly, the presence of a sufficient number of acquisition stars (with observed magnitudes $13.5<H<17$) 
within the patrol field of a given KMOS pointing and preferentially and equally distributed among the 24 arms and three spectrometers/detectors (these stars are used to align KMOS). The second requirement is the absence of bright stars (which would create persistency) superposed with the path of the KMOS arms on the field of view. The final requirement is the presence of at least one bright guide star (with an observed magnitude $9 < R < 12$) in the vicinity of the pointing to maintain telescope tracking. All the aforementioned stellar sources must have low proper motion. Specifically, we required $\mid \mu_{\rm RA} \mid \, {\rm and} \, \mid \mu_{\rm Dec} \mid < 20 {\rm \, mas \, yr^{-1}}$.

The observations cover four distinct fields whose observability spreads adequately throughout the year.
The number of hours allocated per semester and per field is reported in Table \ref{Tab:time_allocated}. The corresponding number of pointings are indicated in parentheses, split between the $H$- and $K$-bands, with a slight preference of $H$-band over $K$-band to maximise the redshift measurement success rate.
A detailed list of the pointings observed in P99--P103 is reported in Table \ref{Tab:observed_pointings}.
Each observing block (OB) is composed of two pointings of 1\,h on sky, which provides about 40 minutes on source. These pointings can either be observed during the same night or on different nights. In the latter case, the observations are reduced separately and then combined.
Only during the last awarded period (P102) was the on-source time for $K$-band pointings doubled in order to increase the detectability of the targeted galaxies.
The data-reduction procedure, described in the next section, is applied to the single science and sky frames separately, and the frames are combined at the end of the reduction, after the whole pointing (two OBs) has been observed.

\subsection{Data reduction}
The data were reduced with the Software Package for Astronomical Reduction with KMOS (SPARK; \citealt{Davies2013}) using recipes outlined in the SPARK instructional guide\footnote{ftp://ftp.eso.org/pub/dfs/pipelines/kmos/kmos-pipeline-cookbook-0.9.pdf}.
The reduction first applies a correction for detector effects, including (1) the correction of the readout channel variations via the reference pixels (pixels without photodiodes but with full electronics readout), and (2) the correction for the picture-frame effects affecting IFUs at the edges of the detector, using median DARK frames. The reduction then proceeds through the standard calibration steps, namely flat fielding, illumination correction, wavelength calibration (the accuracy of the wavelength solution is to the order of $30 \,\si{\kms}$), reduction of the spectrophotometric standards, and finally the data cube reconstruction. After this stage, an additional custom processing was performed on these reconstructed data cubes to further subtract the sky lines. The custom-made sky-line correction routine is an adaptation of the Zurich Atmosphere Purge (ZAP; \citealt{Soto2017}) approach to the KMOS data. The routine subtracts the closest sky frame to the science frame in the O-S-O-O-S-O sequence and then further optimises the fitting to the OH sky-line residuals via a ZAP principal-component analysis \citep{Wisnioski2019}. The background continuum is removed using offset sky frames without attempting to correct for short time scale background variations, and thus some residual continuum levels are still expected. An illumination correction is then applied to flatten out the IFU spatial response. A heliocentric correction is finally performed before the data cubes are combined.

A further set of reduction steps is applied by means of a routine developed by the KMOS GTO team in order to perform the flux calibration and a refined background subtraction \citep{Wisnioski2019}. 
The flux calibration procedure can be summarised in three operations: a) correction for the grism+detector wavelength response using a telluric star; b) application of the zero point to convert fluxes to units of $\rm 10^{-17}\, W \, m^{-2} \, \micron^{-1}$ (to be further multiplied by 0.1 to obtain $\rm erg \, cm^{-2} \, s^{-1}$\,\AA$^{-1}$); and c) fit of the PSF star in the science data with a Moffat function for
the monitoring of the flux and estimation of the PSF from its average FWHM across the frames, and measured again on the combined data cubes for consistency checks.
Individual frames are then median-combined into final cubes using spatial shifts measured from the average centre of the stars within the same pointings (when applicable) or using the information given in the header of each frame.
Variations in flux and seeing among the combined frames are typically 10\% and 0\farcs1, respectively.
A detailed description of the data reduction for KMOS data cubes can be found in \cite{Wisnioski2019}.

\onecolumn
\begin{table*}
\centering
\caption{List of the awarded time (in h) for KMOS observations. Below the number of hours, in parenthesis, the number of the observed pointings is indicated, together with the selected filter, for example, 3$H$+2$K$ means that three pointings have been observed in the $H$-band and two pointings have been observed in the $K$-band.  \label{Tab:time_allocated}}
\begin{tabular}{cccccc}
\hline
Field & P99 & P100 & P101 & P102 & Total\\
\hline
COSMOS & 7.6 & 10.8 & 0 & 10.8 & 29.2 \\
& (2$H$+1.5$YJ^{\star}$) & (3$H$ + 2$K$) & & (5$H$) & (10$H$+2$K$) \\
\hline
ECDFS & 0 & 0 & 2.2 & 0 & 2.2 \\
& & & (1$H$) & & (1$H$)\\
\hline
SXDF & 0 & 8.7 & 5.4 & 10.8 & 24.9 \\
& & (2$H$+2$K$) & (1$H$+1$K$) & (3$H^{\star\star}$+1$K$) & (6$H$+4$K$)\\
\hline
VVDS & 6.5 & 10.8 & 6.5 & 8.7 & 32.5 \\
& (2$H$+1$K$) & (3$H$+2$K$) & (2$H$+1$K$) & (2$H$+2$K$) & (9$H$+6$K$)\\
\hline
Total & 14.1 & 30.3 & 14.1 & 30.3 & 88.8 \\
\end{tabular}\\
$^{\star}${\scriptsize We had initially planned to target sources with $1.8<z_{\rm phot}<2.0$, for which the O{\sc ii} doublet is in the $YJ$-grating. The detection of O{\sc ii} is challenging in high-redshift galaxies, and our first observations in P99 had a low success rate. We therefore decided to start in P100 to exclusively concentrate on the detection of ${\rm H}\alpha$ in the $H$- and $K$-gratings.}\\
$^{\star\star}${\scriptsize The observation of the last three $H$-band pointings in the SXDF field (see Table \ref{Tab:observed_pointings} for details) was carried over P103.}\\
\end{table*}

\begin{longtable}{lccccccc}
\caption{List of the observed pointings.} \label{Tab:observed_pointings}\\
\hline
Pointing ID & RA$_{cen}$ & Dec$_{cen}$ & Exp\_time & Filter & UT Date & Success Rate \\
& (deg) & (deg) & (s) & & (yyyy.mm.dd) & ($3\leq Q \leq 4$/Q=2/Observed)\\ \hline

P99\_COSMOS\_HaHP1 & 149.8900 & 1.9003 & 2 $\times$ 1800 & $H$ & 2017.04.03 & 14/4/22\\
P99\_COSMOS\_HaHP3 & 150.1672 & 1.8391 & 1800 & $H$ & 2017.04.02 & 16/4/22\\
 & & & 1800 & $H$ & 2017.04.04 &\\
P99\_VVDS\_HaHP2 & 36.3758 & $-4.2529$ & 2 $\times$ 1800 & $H$ & 2017.12.23 & 18/2/22\\
P99\_VVDS\_HaHP3 & 36.2548 & $-4.4108$ & 1800 & $H$ & 2017.09.06 & 14/3/22\\
& & & 1800 & $H$ & 2017.09.14 &\\
P99\_VVDS\_HaKP1 & 36.2005 & $-4.0997$ & 1800 & $K$ & 2017.07.12 & 5/5/22\\
 & & & 1800 & $K$ & 2017.09.14 & \\
\hline
P100\_COSMOS\_HaHP1 & 150.3757 & 2.5168 & 1800 & $H$ & 2018.03.03 & 12/1/22\\
 & & & 1800 & $H$ & 2018.03.17 &\\
P100\_COSMOS\_HaHP2 & 150.3964 & 2.4168 & 2 $\times$ 1800 & $H$ & 2018.03.24 & 11/0/22\\
P100\_COSMOS\_HaHP3 & 150.3342 & 2.3114 & 2 $\times$ 1800 & $H$ & 2018.04.07 & 17/2/22\\
P100\_COSMOS\_HaKP1 & 150.3758 & 2.5113 & 1800 & $K$ & 2018.03.17 & 7/1/22\\
& & & 1800 & $K$ & 2018.03.24 &\\
P100\_COSMOS\_HaKP2 & 150.4966 & 2.5003 & 1800 & $K$ & 2018.04.04 & 9/0/22\\
& & & 1800 & $K$ & 2018.04.06 &\\
P100\_SXDF\_haHP1 & 34.6131 & $-5.3581$ & 2 $\times$ 1800 & $H$ & 2017.10.01 & 8/2/22\\
P100\_SXDF\_haHP2 & 34.7924 & $-4.8665$ & 2 $\times$ 1800 & $H$ & 2017.12.25 & 16/0/22\\
P100\_SXDF\_haKP1 & 34.6130 & $-5.3587$ & 1800 & $K$ & 2017.10.26 & 3/1/22\\
 & & & 1800 & $H$ & 2018.07.29 &\\
P100\_SXDF\_haKP2 & 34.7925 & $-4.8669$ & 1800 & $K$ & 2018.07.27 & 1/0/22\\
 & & & 1800 & $H$ & 2018.09.09 &\\
P100\_VVDS\_haHP1 & 36.5006 & $-4.0833$ & 1800 & $H$ & 2018.01.20 & 14/3/22\\
 & & & 1800 & $H$ & 2018.01.21 &\\
P100\_VVDS\_haHP2 &  36.6674 & $-4.4833$ & 1800 & $H$ & 2018.07.29 & 12/5/22\\
 & & & 1800 & $H$ & 2018.08.27 &\\
P100\_VVDS\_haKP1 & 36.6257 & $-4.4940$ & 1800 & $K$ & 2017.11.09 & 7/3/22\\
 & & & 1800 & $K$ & 2018.01.15 &\\
P100\_VVDS\_haKP2 & 36.7924 & $-4.5337$ & 1800 & $K$ & 2018.09.08 & 11/0/22\\
 & & & 1800 & $K$ & 2018.09.09 &\\
P100\_VVDS\_haHP3 & 36.7922 & $-4.4501$ & 1800 & $H$ & 2018.10.30 & 11/0/22\\
 & & & 1800 & $H$ & 2018.10.31 &\\

\hline
P101\_ECDFS\_haHP1 & 53.0840 & $-27.7418$ & 1800 & $H$ & 2018.07.03 & 12/1/22\\
 & & & 1800 & $H$ & 2018.08.30 &\\
P101\_SXDF\_haHP1 & 34.0842 & $-5.1167$ & 1800 & $H$ & 2018.09.03 & 12/0/22\\
& & & 1800 & $H$ & 2018.10.31 &\\
P101\_SXDF\_haKP1 & 34.3047 & $-5.3420$ & 2 $\times$ 1800 & $K$ & 2018.12.09 & 5/1/22\\
& & & 1800 & $K$ & 2018.12.11 &\\
& & & 1800 & $K$ & 2018.12.14 &\\
P101\_VVDS\_haHP1 & 36.8047 & $-4.1669$ & 2 $\times$ 1800 & $H$ & 2018.09.04 & 17/0/22\\
P101\_VVDS\_haHP2 & 36.9217 & $-4.5527$ & 2 $\times$ 1800 & $H$ & 2018.12.14 & 15/1/22\\
P101\_VVDS\_haKP1 & 36.7296 & $-4.4668$ & 2 $\times$ 1800 & $K$ & 2018.11.12 & 8/0/22\\

\hline
P102\_P100\_VVDS\_HaKP1$^{\star}$ & 36.6257 & $-4.4944$ & 1800 & $K$ & 2018.12.20 & 15/3/22\\
& & & 1800 & $K$ & 2018.12.21 &\\
P102\_P99\_VVDS\_HaKP1$^{\star}$ & 36.2009 & $-4.1000$ & 1800 & $K$ & 2018.12.21 & 15/0/22\\
& & & 1800 & $K$ & 2018.12.22 &\\
P102\_VVDS\_HaHP1 & 36.5424 & $-4.8001$ & 1800 & $H$ & 2018.12.22 & 12/3/22\\
& & & 1800 & $H$ & 2018.12.24 &\\
P102\_VVDS\_HaHP2 & 36.3672 & $-4.2446$ & 2 $\times$ 1800 & $H$ & 2018.12.24 & 14/2/22\\

P102\_COSMOS\_HaHP1 & 150.0840 & 2.2193 & 1800 & $H$ & 2019.02.14 & 8/5/22\\
& & & 1800 & $H$ & 2019.02.23 &\\ 
P102\_COSMOS\_HaHP2 & 150.2464 & 1.8080 & 1800 & $H$ & 2019.02.21 & 12/0/22\\
& & & 1800 & $H$ & 2019.02.23 &\\  
P102\_COSMOS\_HaHP3 & 149.7305 & 2.1500 & 2 $\times$ 1800 & $H$ & 2019.02.22 & 12/0/22\\
P102\_COSMOS\_HaHP4 & 149.8884 & 2.5663 & 1800 & $H$ & 2019.02.27 & 14/0/22\\
& & & 1800 & $H$ & 2019.03.12 &\\ 
P102\_COSMOS\_HaHP5 & 150.4503 & 2.0366 & 2 $\times$ 1800 & $H$ & 2019.01.19 & 12/0/22\\
P102\_SXDF\_HaKP1  & 34.6756 & $-5.2782$ & 1800 & $K$ & 2019.01.25 & 1/0/22\\
& & & 1800 & $K$ & 2019.02.13 &\\
& & & 1800 & $K$ & 2019.02.14 &\\
& & & 1800 & $K$ & 2019.02.18 &\\
P102\_SXDF\_HaHP1 & 34.6673 & $-5.2670$ & 1800 & $H$ & 2019.02.19 & 12/3/22\\
& & & 1800 & $H$ & 2019.07.14 &\\
P102\_SXDF\_HaHP2 & 34.2004 & $-5.2056$ & 1800 & $H$ & 2019.07.17 & 9/3/22\\
& & & 1800 & $H$ & 2019.07.18 &\\
P102\_SXDF\_HaHP3 & 34.6981 & $-5.0032$ & 1800 & $H$ & 2019.07.30 & 10/0/22\\
\hline
\end{longtable}
{$^{\star}${\scriptsize These pointings are replicated configurations of two  $K$-band VVDS pointings with low success rates observed during P99 (P102\_P99\_VVDS\_HaKP1) and P100 (P102\_P100\_VVDS\_HaKP1); the overall configuration is maintained, but new objects have been allocated to arms in which a good spectroscopic redshift was derived during the earlier observations (quality flag from three to four, which means that we replaced five to seven galaxies per pointing).}}

\twocolumn

\section{Redshift assignment}
\label{Sec:redshift_assignment}

The observational programme performed with KMOS\MVAt VLT aims to derive the spectroscopic redshift of $1.3\lesssim z_{\rm phot} \lesssim 2.5$ galaxies through a single emission line, mainly ${\rm H}\alpha$ in the $H$- and $K$-band filters.

Each observed spectrum was analysed by two co-authors to independently determine the redshift and the quality flag. The results were then reconciled and discussed by the two people.
We developed an interactive routine that we applied to the reduced and combined data cubes for the redshift assignment. There are several steps towards the application of the code:
\begin{itemize}
\item when continuum is visible, find the position of the targeted source in the spatial plane of the median image of the data cube, otherwise we use the nominal centre at the pixel with coordinates $(x,y)=(9,9)$;
\item create two-dimensional (2D) vertical/horizontal spectra computing the median flux at each wavelength of four lines/columns around the central pixel;
\item identify the presence of an emission line either in the vertical and/or in the horizontal 2D spectrum and select a narrower (about 10 pixels) wavelength range to determine the pixels where the emission is detected;
\item plot the $(x,y)$ spatial image of the cube at four pixels corresponding to the wavelengths where the emission has the highest intensity in order to identify both the wavelength (in pixel units) of the peak of the emission and the $(x,y)$ coordinates of its centre;
\item plot the 1D spectrum of the selected central spaxel and the 1D spectrum obtained by summing the flux in a number of adjacent pixels to increase the signal to noise (the number of pixels varies from a cross of five to a square of nine, depending on the spatial extension of the source);
\item perform a Gaussian fit weighted by the noise spectrum on the identified emission line;
\item choose the most appropriate-looking value of the emission-line centre, between the position of the mean of the fitted Gaussian and the position of the peak pixel;
\item compute the redshift with the formula
\begin{equation}
z_{\rm spec}=(\lambda_{\rm peak/Gaussian}-\lambda_{{\rm H}\alpha})/\lambda_{{\rm H}\alpha}, 
\end{equation}
where $\lambda_{\rm peak/Gaussian}$ is the wavelength (in \micron) corresponding to the pixel peak or to the centre of the fitted Gaussian, and $\lambda_{{\rm H}\alpha}$ is the H$\alpha$ vacuum wavelength expressed in \micron.
\end{itemize}

\subsection{Quality flags}
Each redshift measurement is assigned a preliminary quality flag reproducing the flagging scheme presented in M17:
\begin{itemize}
    \item $Q=4$: indicates a secure redshift measurement based on the identification of more than one emission line. Specifically, the H$\alpha$ line is associated with the N{\sc ii} doublet at $\lambda$6549.84\,\AA, $\lambda$\,6585.23\,\AA. In one case, the O{\sc ii} doublet ($\lambda$\,3727.09\,\AA~and $\lambda$\,3729.88\,\AA) was identified rather than the H$\alpha$ line. (Details on how the identification and fit of these groups of lines is performed is given in Sect. \ref{subsec:Halpha_kubeviz});
    \item $Q=3.5$: indicates a secure redshift measurement based on a single emission line (usually H$\alpha$);
    \item $Q=3$: indicates a likely secure redshift determination, but with a low probability of an incorrect identification or an uncertain redshift due to low signal-to-noise data or sky-line contamination affecting the Gaussian fit;
    \item $Q=2$: flag 2 indicates a reasonable but not secure enough guess. The targets being assigned with this flag are discarded from the calibration sample, and not included in the released catalogue.
\end{itemize}

\subsection{Refine the redshift assignment with \texttt{KUBEVIZ}}
\label{subsec:Halpha_kubeviz}

Maps of the emission-line fluxes were obtained from the reduced data cube using the IDL routine \texttt{KUBEVIZ} \citep{Fossati2016}.
The code simultaneously fits groups of lines (defined as `line sets', e.g. H$\alpha$ and the N{\sc ii} $\lambda$6548.05, $\lambda$6583.45 doublet, or the O{\sc iii} $\lambda$\,4958.91, $\lambda$\,5006.84 doublet) using a combination of 1D Gaussian functions with fixed relative velocities.
The continuum level is evaluated as the median value of the flux with an intensity from 40\% to 60\% within the total range of values inside two symmetric wavelength windows around each line set, and then subtracted.
During the fit, KUBEVIZ takes into account the noise from the `stat' data cube, thus optimally suppressing sky-line residuals. Furthermore, we reject the spaxels with ${\rm SNR}<4.0$ from the fit, and manually reject bad-fit and isolated spaxels from the map.

There are several aspects that motivated us to use \texttt{KUBEVIZ} on the KMOS reduced data cubes. Firstly, fitting the H$\alpha$+N{\sc ii} lineset improves the $z_{\rm spec}$ measurement; starting from the H$\alpha$ emission map of the galaxy and its corresponding velocity ($v$) map, we arbitrarily chose the centre ($v=0$) of the galaxy as the spaxel that best compromises the peak of the H$\alpha$ emission with the centre of the galaxy signal/velocity map (if present), and we corrected the input $z_{\rm spec}$ and the relative velocity of every spaxel accordingly.
Furthermore, a successful \texttt{KUBEVIZ} fit of low-quality spectroscopic candidates (those that were assigned a $Q=2$ flag at the redshift assignment stage) allows their spectroscopic confirmation by promoting the quality flag of the $z_{\rm spec}$ measurement, and thus their inclusion in the calibration sample.
Finally, the \texttt{KUBEVIZ} outputs constitute the groundwork for measuring the total H$\alpha$ flux of the sources, which is described in detail in Sect. \ref{Sec:Halpha_flux}.

\subsection{Collecting multi-band photometry}
\label{Sec:photometric_catalogues}

We collected all available multi-wavelength photometry for the galaxy sample observed during the KMOS programme from public data releases in the three fields \footnote{The multicolour photometry used here is optimised to measured physical parameters of galaxies of known spectroscopic redshift;
other choices might be preferable when computing photometric redshifts (see \citealt{Masters2015}).}.

\subsubsection{COSMOS}
\label{subsec:COSMOS_photometry}
We start from the COSMOS2015 catalogue released in \cite{Laigle2016}, which contains precise PSF-matched photometry for more than half a million sources in the COSMOS field.
Among the wide collection of photometric bands available in the data release, we selected CFHT $u^\prime$ and Subaru $B, V, R, i^+, z^+$ and $z^{++}$ optical aperture magnitudes (3\arcsec),  $Y J H Ks$ near-infrared aperture magnitudes (3\arcsec) from the UltraVISTA-DR2 survey, mid-infrared data from the {\it Spitzer} Large Area Survey with Hyper-Suprime-Cam (SPLASH) legacy programme (IRAC ch1, ch2, ch3, ch4 total magnitudes), and GALEX $NUV$ total magnitudes.

We computed total magnitudes in the optical and near-infrared domain starting from the aperture magnitudes and the correction factors given in the released catalogue using Eq. (9) in the Appendix of \cite{Laigle2016}:
\begin{equation}
\label{Eq:mag_aper_to_tot}
    {\tt MAG\_TOTAL}_{i,f}={\tt MAG\_APER3}_{i,f} + o_i - s_f,
\end{equation}
where $i$ identifies the single objects, $f$ the considered filter, {\tt MAG\_APER3} is the magnitudes computed within a 3\arcsec radius aperture contained in the catalogue, $o_i$ is the photometric offset computed for scaling aperture magnitudes to total ones, and $s_f$ is the systematic offset computed in the paper using spectroscopic redshifts.
Finally, all magnitudes should also be corrected for foreground Galactic extinction using the reddening values given in the released catalogue for each object (Eq. 10 in the Appendix):
\begin{equation}
\label{Eq:mag_tot_ext_corr}
    {\tt MAG\_TOTAL}_{i,f, {\rm extcorr}}={\tt MAG\_TOTAL}_{i,f} - { E(B-V)}_i \times F_f ,
\end{equation}
where $F_f$ is the extinction factor of any given filter.

Besides the photometric information, we also kept the $z_{\rm phot}$ and physical properties ($E(B-V)$, absolute magnitudes, median stellar masses, and SFR from the maximum likelihood -- ML -- analysis of \texttt{LePhare}) derived in \cite{Laigle2016} by means of the SED fitting code \texttt{LePhare} \citep{Arnouts1999,Ilbert2006} run on the complete 30-band photometric data set.

\subsubsection{SXDF}
\label{subsec:SXDF_photometry}

We collected multi-band photometry in the SPLASH survey data release \cite{Mehta2018}.
We considered optical aperture magnitudes (3\arcsec) from CFHT $u$ filter and from the Hyper Suprime-Cam (HSC) UltraDeep layer in the $griz$ filters; the near-infrared regime is fully covered by the VISTA Deep Extragalactic Observations (VIDEO) Survey $Y J H Ks$ aperture magnitudes (3\arcsec), and the mid-infrared takes advantage of the IRAC coverage (ch1, ch2, ch3, ch4) from SPLASH.

Aperture magnitudes were corrected to total values using the offsets given in the released catalogue table ({\tt OFFSET\_MAG}) and all magnitudes were corrected for foreground extinction following the same procedure described in Sect. \ref{subsec:COSMOS_photometry} for the COSMOS field.
Consistent with what was done in \cite{Laigle2016} for the COSMOS field, \cite{Mehta2018} performed the SED fitting analysis of the SXDF photometric sample using \texttt{LePhare}. We took advantage of the outputs of their analysis to collect the physical properties of all our observed galaxies ($E(B-V)$, absolute magnitudes, best fit stellar masses, and SFRs).

\subsubsection{VVDS}
\label{subsec:VVDS_photometry}

A complete and homogeneous collection of photometry in the VVDS-02h field is contained in the VIDEO Survey, which has been merged with the CFHTLS Deep1 optical ($ugriz$) catalogue (Jarvis, M. \& H{\"a}ussler, B., priv. comm.).
The catalogue contains aperture magnitudes within a $2\arcsec$ radius measured in a homogeneous manner in all the optical and near-infrared filters.
We computed the aperture to total magnitude offsets using the \texttt{SExtractor} MAG\_AUTO values given in the catalogues and the photometric errors, according to Eq. (4) and (5) in \cite{Laigle2016}:
\begin{equation}
\label{Eq:mag_auto_total_offset}
    o=\frac{1}{\sum_{{\rm filters} \, i} w_i} \times \sum_{{\rm filters} \, i} ({\tt MAG_{AUTO}}-{\tt MAG_{APER}})_i \times w_i
,\end{equation}
where
\begin{equation}
\label{Eq:weights_mag_auto_aper}
    w_i=\frac{1}{(\sigma^2_{\tt AUTO} + \sigma^2_{\tt APER})_i}.\end{equation}
The offsets are computed for each object in the catalogue ($i$) using all the bandpasses in the optical and near-infrared domain.
We finally corrected total magnitudes for Milky Way foreground extinction using the \cite{Schlegel1998} maps (consistent with what was used in \citealt{Laigle2016}) at the coordinates of each object  and using the appropriate filter factors, as given in Eq. (\ref{Eq:mag_tot_ext_corr}).

In order to investigate and compare the properties of all the observed galaxies with the spectroscopically confirmed ones, and to have consistent $z_{\rm phot}$ measurements throughout the three explored fields, we ran \texttt{LePhare} on the whole set of collected filters and derived $z_{\rm phot}$ and physical properties of all observed VVDS galaxies ($E(B-V)$, absolute magnitudes, median stellar masses, and SFR from the ML analysis).

\section{Results I: The success rate of the redshift assignment}
\label{Sec:results_I}

In light of the concepts outlined above, the success rate (SR) of the KMOS spectroscopic campaign in the context of the C3R2 survey must be evaluated in two ways: (1) as any spectroscopic survey, as the ratio (or, equivalently, percentage) of the total number of high-quality $z_{\rm spec}$ measured with respect to the number of targets observed; (2) as the total number of empty/undersampled cells that are newly filled with spectroscopically confirmed galaxies. Needless to say, these two quantities should be considered together: a large number of high-quality $z_{\rm spec}$ assigned to a small number of cells is less valuable than a smaller number of high-quality $z_{\rm spec}$ covering a larger number of empty SOM cells.

The total number of $z_{\rm phot}$ targets observed with KMOS was 805, 424 of which provided a secure redshift measurement ($ Q\geq 3$), leading to a total SR of 51.4\%.
The detailed SR of the four semesters and two filters is listed Table \ref{Tab:SR}.
Overall, the SR of $H$-band observations is twice that of the $K$-band observations, likely primarily due to the higher backgrounds at longer wavelength. Additional challenges are caused by the lowering of the precision of currently available template fitting techniques as redshift increases, and also the lower brightness of the targets themselves. Doubling the exposure time of $K$-band pointings and repeating the observation of two $K$-band pointings observed during P99 and P100, was not conclusive in this respect: the $K$-band SR in P102 only slightly increased compared to previous periods. Whether this result is mainly due to the limited accuracy of $z_{\rm phot}$-based target selection or to the necessity of longer exposure times to increase the SNR of the spectra is still unclear, but a detailed analysis of the spectroscopic failures is presented in Sect. \ref{subsec:spec_failures}. 

\begin{table*}
\centering
\caption{Success rate of KMOS observations. \label{Tab:SR}}
\begin{tabular}{ccc}
\hline
Period & $H$-band & $K$-band \\
\hline
P99 & 72/88$^{\star}$ (81.8\%) & 5/22$^{\star\star}$ (22.7\%) \\
P100 & 106/176 (60.2\%) & 46/132 (34.8\%) \\
P101 & 53/89 (59.6\%) & 13/44 (29.5\%) \\
P102 & 117/220 (53.2\%) & 12/51 (30.4\%) \\
Total & 348/573 (60.7\%) & 76/232 (32.8\%) \\
\hline
\end{tabular}\\
$^{\star}${\scriptsize 72 galaxies with accurate $z_{\rm spec}$ estimate ($\rm Q\geq 3$) over 88 observed targets.}\\
$^{\star\star}${\scriptsize Pointing re-observed during P102. Since 17 out of 22 galaxies were re-observed, the contribution to the total number of observed objects in the $K$-band from P99 is just five.}\\
\end{table*}

\begin{figure*}
    \centering
    \includegraphics[scale=0.4]{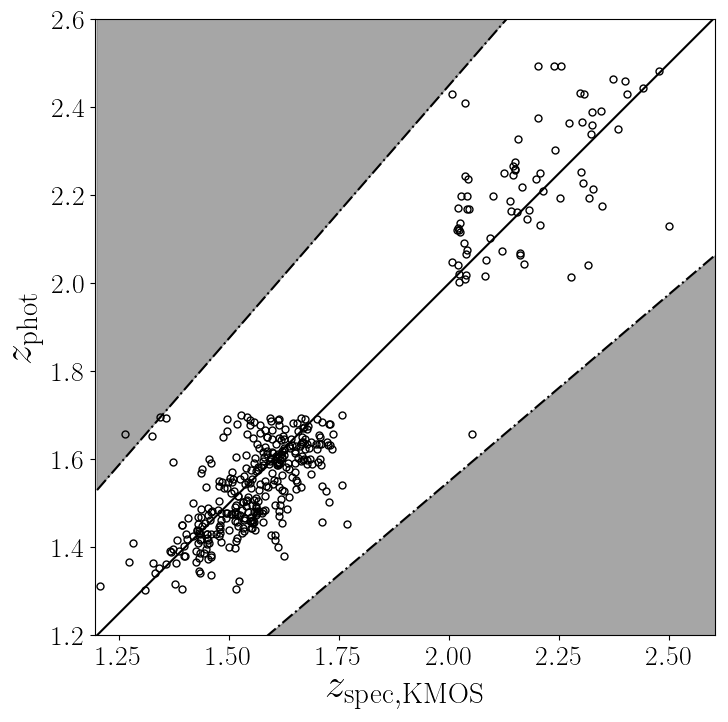}
    \includegraphics[scale=0.4]{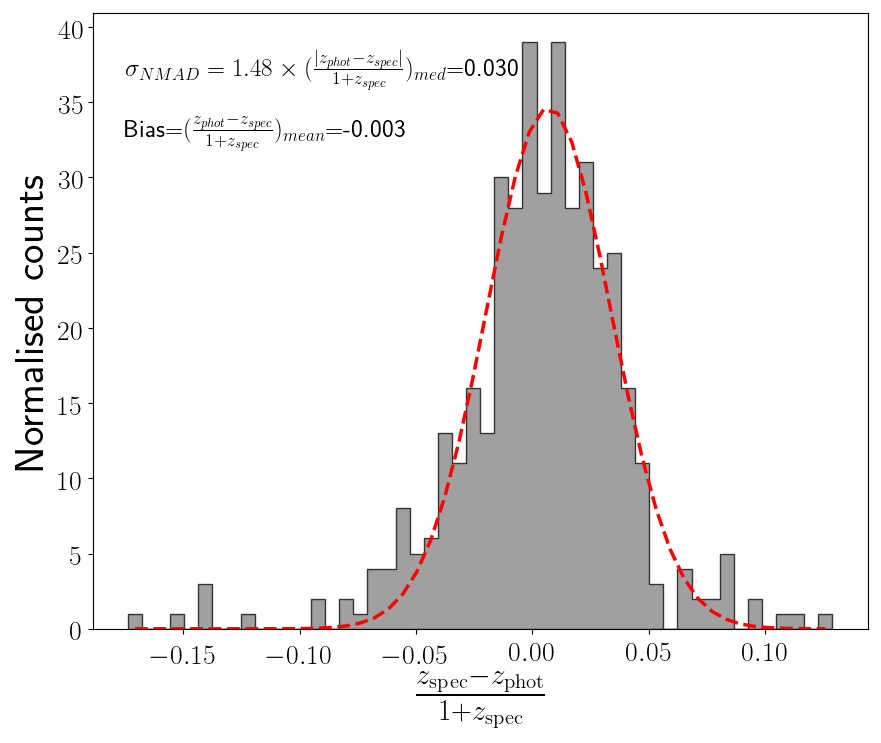}
    \includegraphics[scale=0.4]{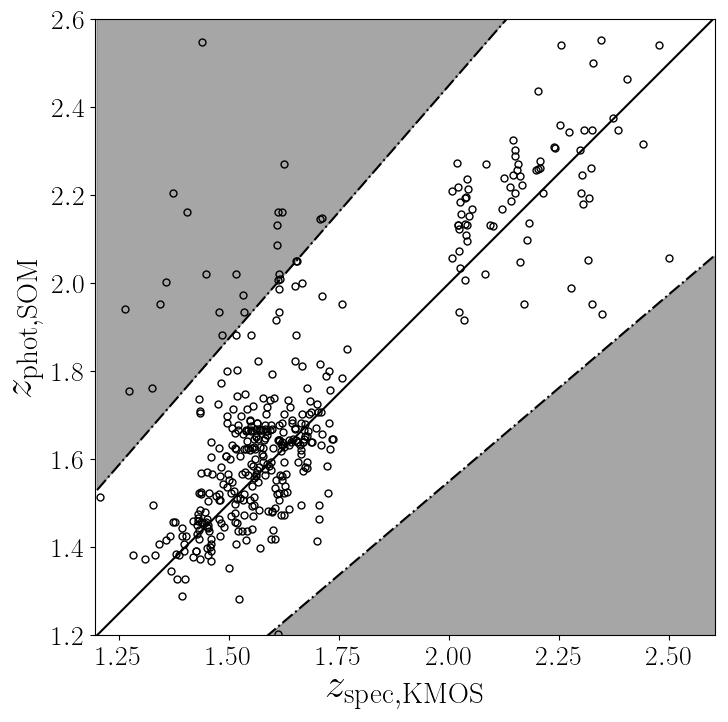}
    \includegraphics[scale=0.4]{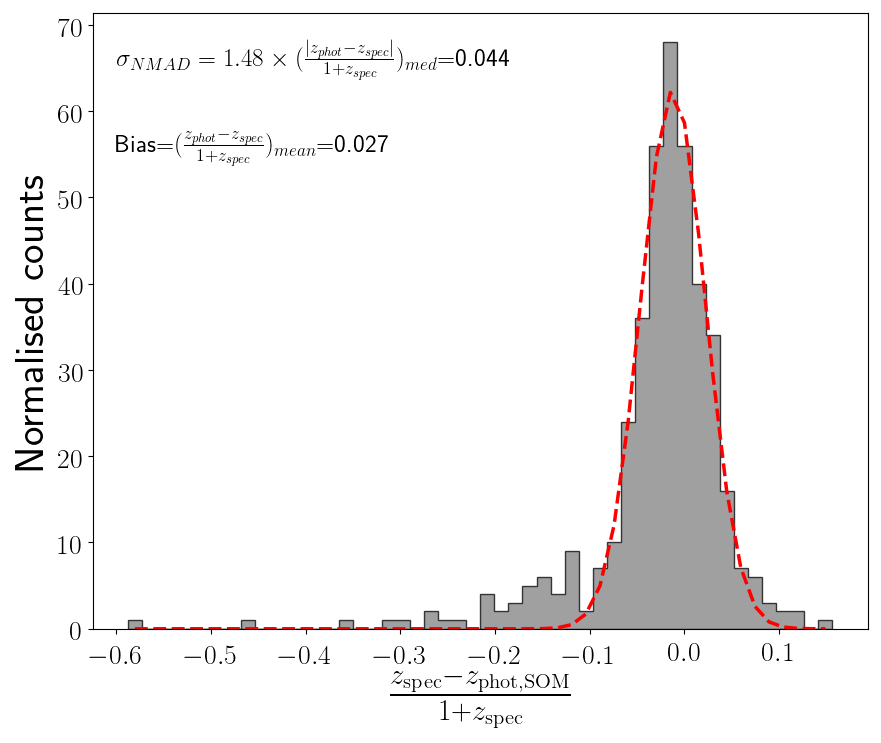}
    \caption{{\it Top left}: comparison between $z_{\rm phot}$ and $z_{\rm spec}$ for high-quality ($Q \geq 3$) redshift galaxies observed during the four periods of the KMOS Large Programme. Lower redshift targets are observed with the $H$-band grism, higher redshift ones with the $K$-band. The dashed lines define the region outside which the $z_{\rm phot}$ is considered a `catastrophic failure' (grey area in the plot), defined by a redshift error $|z_{\rm phot}-z_{\rm spec}|/(1+z_{\rm spec}) \geq 15\%$. {\it Top right}: histogram of the $(z_{\rm phot}-z_{\rm spec})/(1+z_{\rm spec})$ of all high-quality redshift targets. A Gaussian with mean and sigma equal to the bias and $\sigma_{\rm NMAD,}$ respectively, is overplotted with a red dashed line.
    {\it Bottom left}: same as the {\it top-left} panel but comparing $z_{\rm phot,SOM}$ and $z_{\rm spec}$. {\it Bottom right}: same as the {\it top-right} panel but with $z_{\rm phot,SOM}$.}
    \label{fig:zspec_vs_zphot}
\end{figure*}

Figure \ref{fig:zspec_vs_zphot} presents a comparison between the photometric (individual and SOM-based) redshifts and high-quality ($Q \geq 3$) KMOS spectroscopic redshifts.
The dashed lines trace the boundaries outside which the photometric redshifts are considered catastrophic outliers, $|z_{\rm phot}-z_{\rm spec}|/(1+z_{\rm spec}) \geq 15\%$.
The top panels of Fig. \ref{fig:zspec_vs_zphot} compare the individual $z_{\rm phot}$ redshift estimates with our $z_{\rm spec}$ measurements: according to these quantities, our sample contains one catastrophic outlier. This galaxy, observed in the $H$-band, has a $z_{\rm phot}=1.6565$, $z_{\rm spec}=1.2632$ and $Q=3.0$. A detailed analysis of this target revealed a discrepancy between the $individual$ (from template fitting) and the SOM-based $z_{\rm phot}$ estimates ($z_{\rm phot,SOM}=1.9407$), which could be the reason of the misplacement of this target in the $z_{\rm spec}$-$z_{\rm phot}$ plane.
Furthermore, we notice that there is a target observed in the $H$-band with $z_{\rm phot}\leq 1.6$, but validated at $z_{\rm spec}\geq 2$, thanks to the identification of the O{\sc iii} ($\lambda$\,4960.30\,\AA, 5008.24\,\AA) lines.
The bottom panels of Fig. \ref{fig:zspec_vs_zphot} show the same statistical analysis to compare the $z_{\rm spec}$ with the redshift of the SOM cell each galaxy belongs to ($z_{\rm phot,SOM}$).

We point out that the SOM is not intended to be used for individual redshift estimates, and therefore one should not be surprised that its performance in terms of recovering individual $z_{\rm phot}$ values is worse than for individual multi-band template fitting. However, comparing the distribution of $z_{\rm phot}$ and $z_{\rm spec}$ in individual SOM cells is fundamental for a better understanding of cell occupation (e.g. in order to quantify the $z_{\rm phot}$ dispersion of galaxies occupying the same cell or to pinpoint multiple peaks in the distribution of galaxies) and for highlighting problematic regions in  the SOM.

\begin{figure*}
    \centering
    \includegraphics[scale=0.45,trim={0.5cm 5cm 1cm 2.5cm 0},clip]{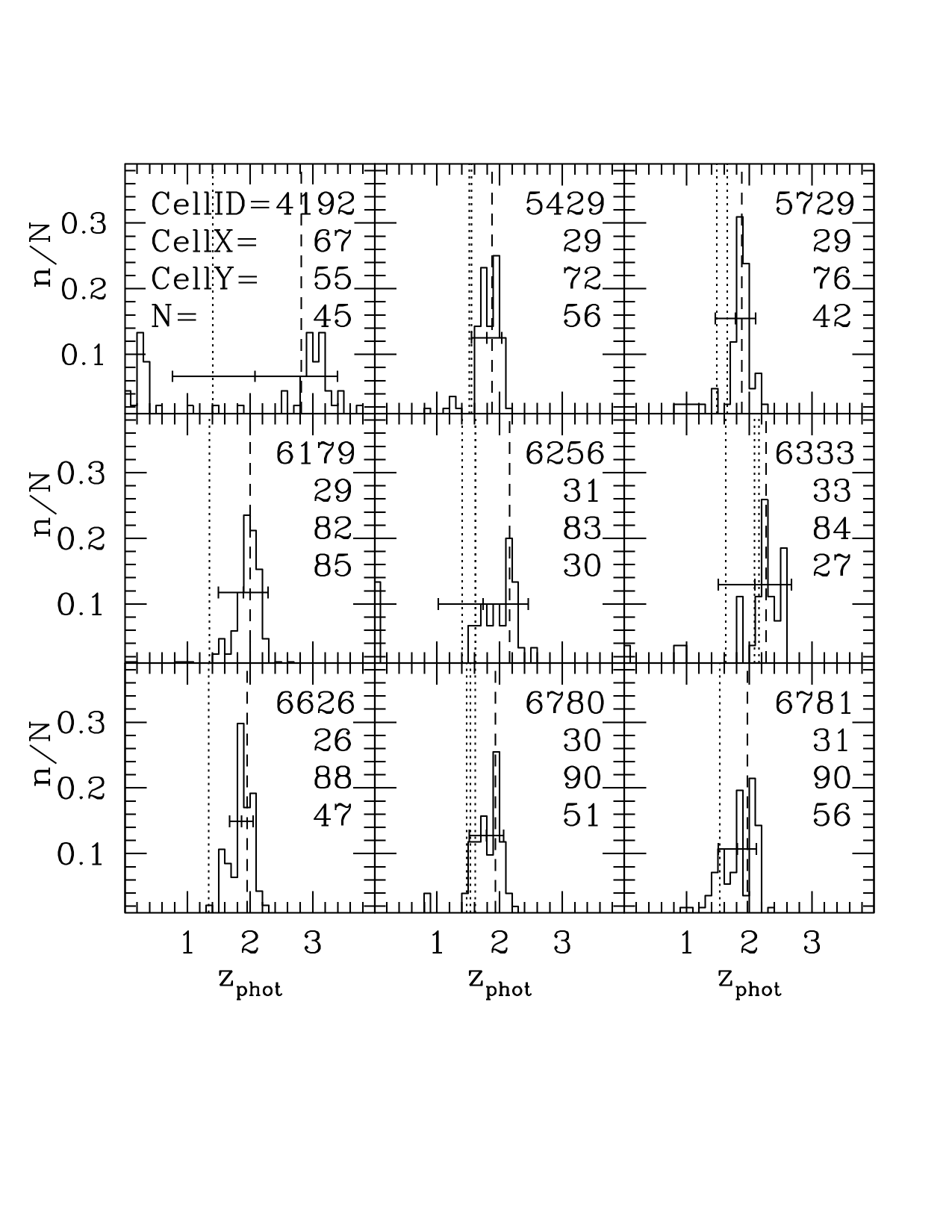}
    \includegraphics[scale=0.45,trim={0.5cm 5cm 1cm 2.5cm 0},clip]{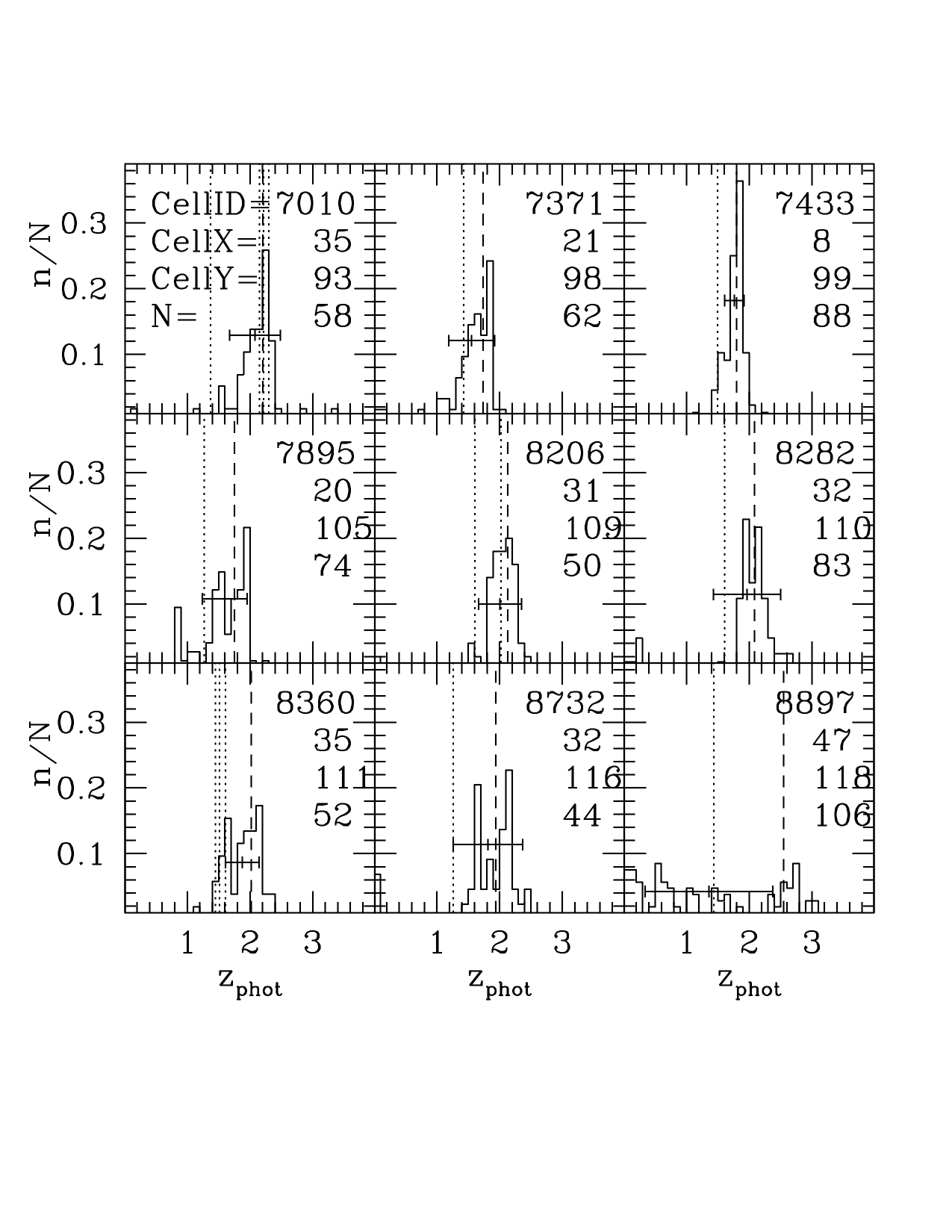}
    \caption{Histogram of $z_{\rm phot}$ of galaxies populating each cell falling in the grey region of the $z_{\rm phot,SOM}$-$z_{\rm spec}$ plane (bottom left panel of Fig. \ref{fig:zspec_vs_zphot}). The distribution is normalised by dividing the number of galaxies in each $z_{\rm phot}$ bin by the total number of $z_{\rm phot}$ populating the considered cell; the number is indicated with the letter N in the top left panel of the figures, and written at the same position in the others. Similarly, the cell number (CellID) and coordinates (CellX, CellY) are also given inside each panel. The $z_{\rm phot,SOM}$ is represented by the dashed line, whereas dotted lines indicate $z_{\rm spec}$ measured during our KMOS programme. The horizontal bar centred on the mean $z_{\rm phot}$ is the rms of the histogram.}
    \label{fig:cell_occupation_zphot}
\end{figure*}

The incidence of catastrophic outliers is significantly higher when $z_{\rm phot,SOM}$ is considered. These 25 galaxies fall into 18 different cells in the SOM, and have an individual $z_{\rm phot}$ more in line with the measured $z_{\rm spec}$; furthermore, in case of multiple observations within the same SOM cell, these galaxies have individual redshifts, which are in line with the other galaxies populating the cell. This result leads us to conclude that there is a misalignment between the redshift of the cell and the redshift of the individual galaxies that compose it.
A better understanding of the distribution of individual $z_{\rm phot}$ of galaxies in the aforementioned SOM cells is given in Fig. \ref{fig:cell_occupation_zphot}. All galaxies in the C3R2 parent $z_{\rm phot}$ sample are used to populate the cells, and the $z_{\rm phot,SOM}$ is also represented inside each panel with the dashed vertical line. As is noticeable from the dispersion values of the histograms (horizontal errorbars centred on the mean $z_{\rm phot}$), the $z_{\rm phot}$ distribution peaks close to the $z_{\rm phot,SOM}$ value, but high dispersion and/or double peaks are present in many of the cells; multiple spectroscopic redshift measurements occupy a narrow redshift range in the panels, often separated from the $z_{\rm phot,SOM}$. Euclid galaxies that are assigned to these problematic cells need to be flagged, as their photometric redshift could be difficult to calibrate. 

The mean value of the redshift difference
\begin{equation}
    \centering
    {\rm mean} \left(\frac{z_{\rm phot} - z_{\rm spec}}{1+z_{\rm spec}}\right)
    \label{eq:bias}
\end{equation}
 is represented as the mean value of the (red dashed) Gaussian in Fig. \ref{fig:zspec_vs_zphot}. When comparing $z_{\rm spec}$ with the individual $z_{\rm phot}$, the value is $-0.0029$, and $-0.0070$ and 0.0148 separately in the $H$- and $K$-bands, respectively, further confirming the decreasing precision of current photometric redshift estimates with increasing redshift. The redshift difference increases to 0.027 when considering the comparison between $z_{\rm spec}$ and $z_{\rm phot,SOM}$, and 0.030, 0.013 in the $H$- and $K$-bands, respectively. The higher $H$-band bias reflects the increased number of catastrophic outliers, which are all located at $z_{\rm spec} \leq 1.75$.

The normalised median absolute deviation, a dispersion measure that is not sensitive to catastrophic outliers \citep{Ilbert2009,Dahlen2013}, defined as
\begin{equation}
    \centering
    \sigma_{\rm NMAD} = 1.48 \times {\rm median} \left(\frac{|z_{\rm phot} - z_{\rm spec}|}{1+z_{\rm spec}}\right)
    \label{eq:sigma_nmad}
,\end{equation}
is 0.0301 (3\%) when individual $z_{\rm phot}$ are considered, and 0.0443 ($\gtrsim 4$\%) when $z_{\rm phot,SOM}$ are used, pointing out that not only the number of catastrophic outliers increases, but also the dispersion of the data points in the white region of the (left-hand panels) scatter plots in Fig. \ref{fig:zspec_vs_zphot}.
The values of the $\Delta \langle z\rangle$ and $\sigma_{\rm NMAD}$ are in agreement with the results presented in M17 and M19.

\begin{figure}
    \centering
    \includegraphics[scale=0.45]{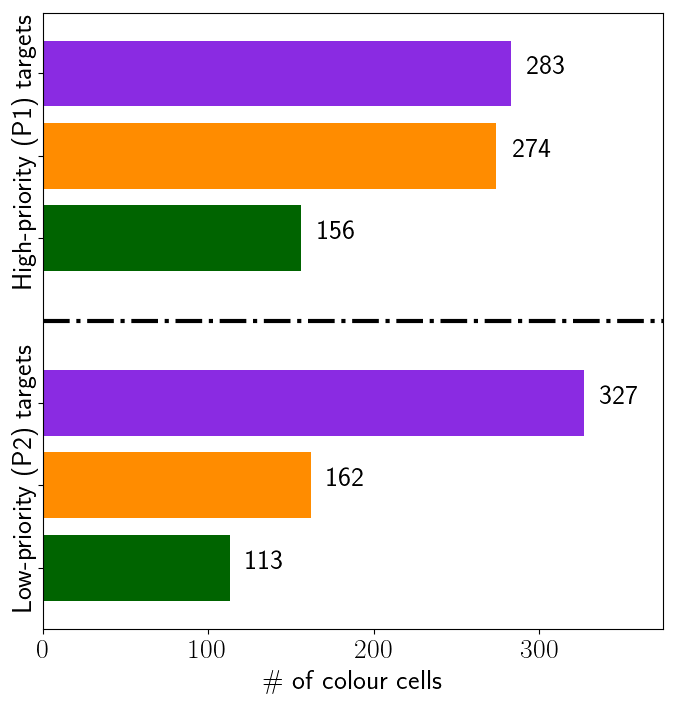}
    \caption{Success rate in terms of number of cells filled with high-quality $z_{\rm spec}$. The observed targets are divided into high (P1) and low (P2) priority targets according to the prioritisation scheme described in Sect. \ref{Sec:prioritization_scheme}. Purple horizontal bars represent the total number of undersampled cells requiring $z_{\rm spec}$ measurements; orange histograms represent the number of cells targeted by all KMOS observations, and green histograms represent the number of cells that provided accurate $z_{\rm spec}$ measurements.}
    \label{fig:SR_cells}
\end{figure}

We computed the number of cells containing P1/P2 targets (according to the priorities defined in Sect. \ref{Sec:prioritization_scheme}) with a SOM photometric redshift $1.3< z_{\rm phot,SOM}<1.7$ (for $H$-band targets) and $2.0< z_{\rm phot,SOM}<2.5$ (for $K$-band targets). The SOM has a number of P1 and P2 cells in this redshift range of 283 and 327, respectively. These numbers indicate the nominal goal of C3R2 in the near-infrared, and will be used as a reference.
The number of P1/P2 cells covered by all KMOS observations (i.e. by all targets placed in KMOS pointings from P99 until P103) is 274 and 162, respectively.
 Of the P1 cells occupied by the KMOS $z_{\rm phot}$ candidates, 57\% (156/274) were spectroscopically confirmed, and the percentage increases to 70\% (113/162) for the P2 targets.
The result is represented in Fig. \ref{fig:SR_cells}.
The histograms shown in Fig. \ref{fig:SR_cells} clearly mirror our observing strategy; we preferentially observed P1 targets covering empty SOM cells, and used P2 targets as fillers for optimising and maximising the number of observed galaxies in one pointing.

\subsection*{Spectroscopic failures and uncalibrated cells}
\label{subsec:spec_failures}

We next analysed the properties of galaxies that were observed but for which we could not assign a spectroscopic redshift.
The main purpose of this analysis is to understand whether there are biases in the data and where these failures are located in the SOM.
To this end, we considered the physical parameters derived from SED fittings in \cite{Laigle2016} and \cite{Mehta2018} for the COSMOS and the SXDF field, respectively. The reason for this choice is twofold. First, when trying to explore the properties of non-spectroscopically validated galaxies, we are forced to rely on $z_{\rm phot-}$ and $z_{\rm phot}$-based physical parameters, which are better determined when a broader photometric sample in terms of the number of available filters is used. Both \cite{Laigle2016} and \cite{Mehta2018} based their SED fitting analyses on a broad number of filters spanning the whole spectrum. Furthermore, the two are comparable as the same PSF homogeneisation was adopted for the data, and the same template library was used for photometric redshift calculation. Secondly, our LePhare setup is a close imitation of what was performed in the two data releases, though limited to a restricted number of filters. In order to check that we did not introduce any bias, we ran LePhare on the photometric samples with the same configuration described in Sect. \ref{Sec:results_II}, but without fixing the redshift, and we compared the results with those from \cite{Laigle2016} and \cite{Mehta2018}. In the COSMOS field, the average difference between stellar masses is 0.090 with an rms of 0.17, and between the (SED fitting based) SFRs it is 0.003 with an rms of 0.229. In the SXDF field, the average difference between stellar masses is 0.069 with a rms of 0.313 and between the (SED fitting based) SFR is 0.237 with a rms of 0.473.
In light of the above, our set of physical parameters is compatible within the errors with the literature but with larger uncertainties. Although all the conclusions discussed below do not change with our derivation, in the following we always refer to the results from the literature.

\begin{figure*}
    \centering
    \includegraphics[scale=0.45]{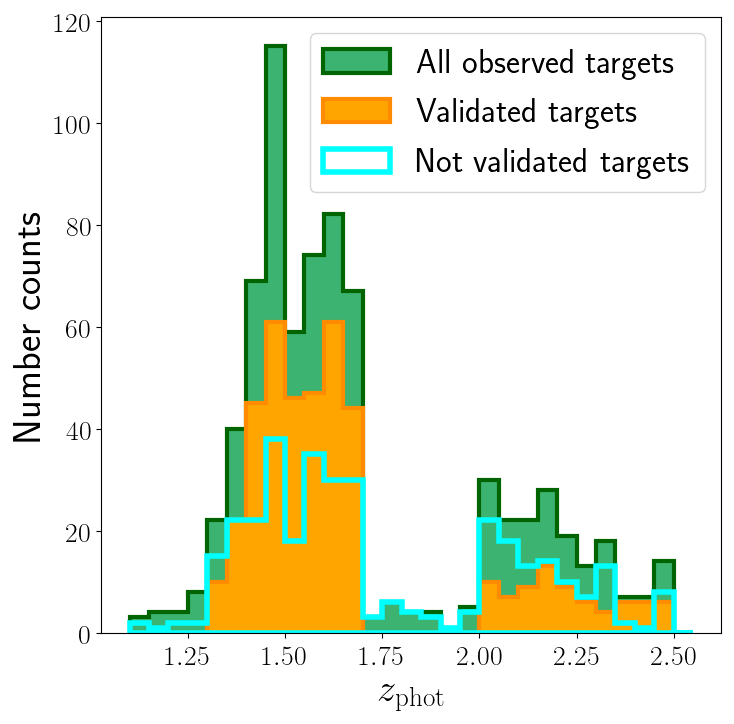}
    \includegraphics[scale=0.45]{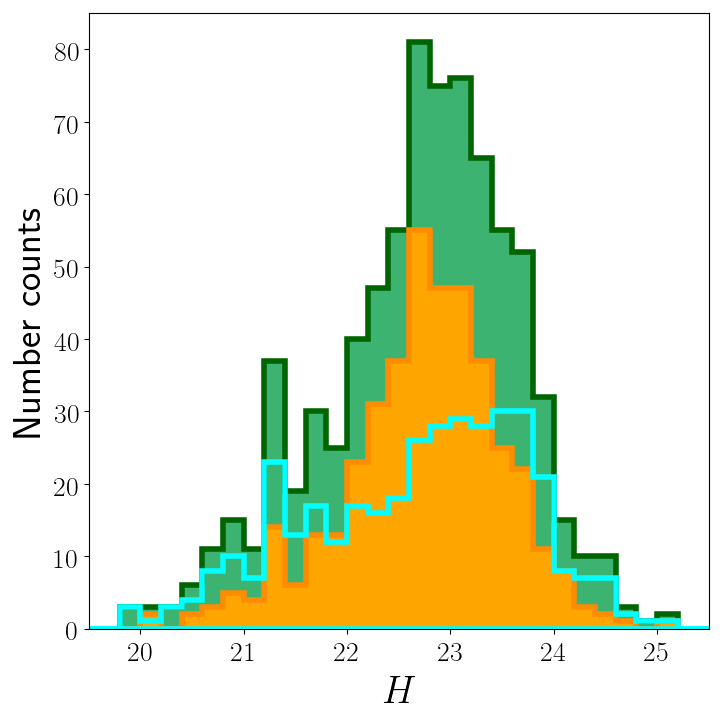}
    \includegraphics[scale=0.45]{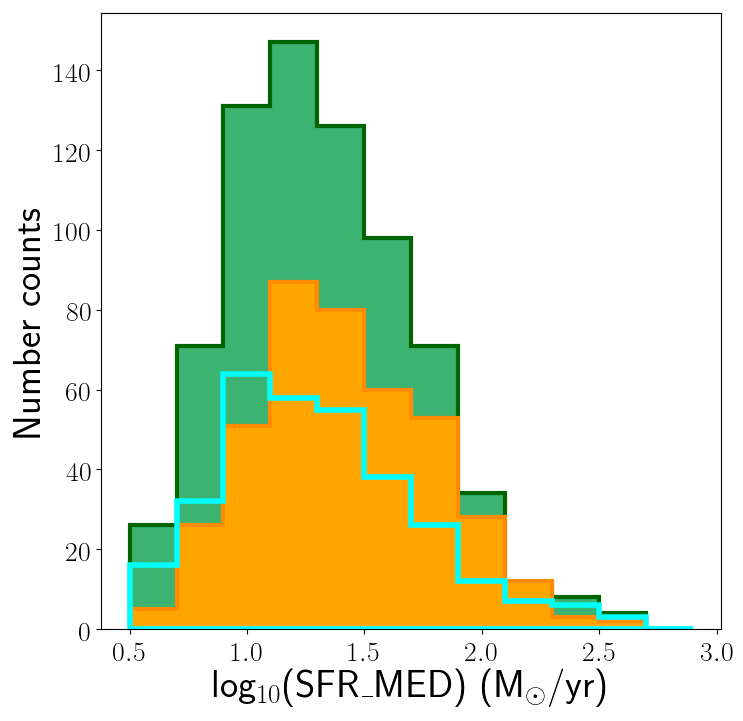}
    \includegraphics[scale=0.45]{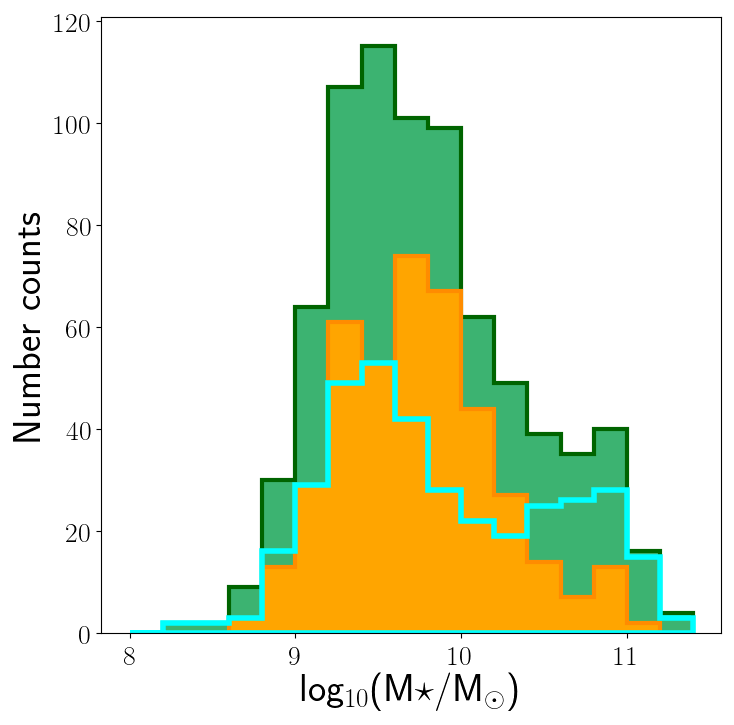}
    \caption{{\it Top Left}: histogram of $z_{\rm phot}$ of individual galaxies from the literature. {\it Top right}: histogram of the observed $H$ total magnitude for all observed targets (green filled), for those with high-quality spectroscopic redshifts (validated targets; orange filled) and for those that could not be assigned a spectroscopic redshift (not validated targets; open blue line). {\it Bottom left}: histogram of the SFR derived from SED fitting for the same samples. {\it Bottom right}: histogram of the stellar mass derived from SED fitting for the same samples.}
    \label{fig:histo_observed_validated_targets}
\end{figure*}

Figure \ref{fig:histo_observed_validated_targets} illustrates the distributions of the $z_{\rm phot}$, observed total $H$ magnitudes and SED-fitting star formation rates (SFRs), and stellar masses for all galaxies observed during our KMOS programme (green histograms), for the sub-samples of spectroscopically confirmed targets (orange histograms) and for the targets that could not be assigned a redshift (blue open histograms).
The distributions of validated and non-validated targets present some differences, with the former being slightly brighter with a higher star formation rate: the median value of $H$ is 22.78 in the validated sample and 22.84 in the non-validated one. Similarly, the median $\logten({\rm SFR}/\rm \si{\solarmass} yr^{-1}$) values are 1.41 and 1.21 in the two samples, respectively. From the bottom right panel of the figure, we can finally notice that our spectroscopic completeness, in terms of number of galaxies validated with respect to the total number of galaxies observed, is a function of stellar mass. Specifically, at low stellar masses ($\logten (M_\star/\si{\solarmass}) < 9.5$), the fraction of validated targets is around 0.5, likely reflecting the low SNR deriving from the limited integration time of our observations; the ratio between validated targets and observed ones reaches the value of 0.7 at $ 9.5 < \logten (M_\star/\si{\solarmass}) < 10$ and finally decreases to the lowest values at higher stellar masses.
A better understanding of the reasons that prevented us from assigning a high-quality spectroscopic redshift to all galaxies can be reached by analysing the distribution of the validated and non-validated targets in the SOM.

\begin{figure*}
    \centering
        \includegraphics[scale=0.275]{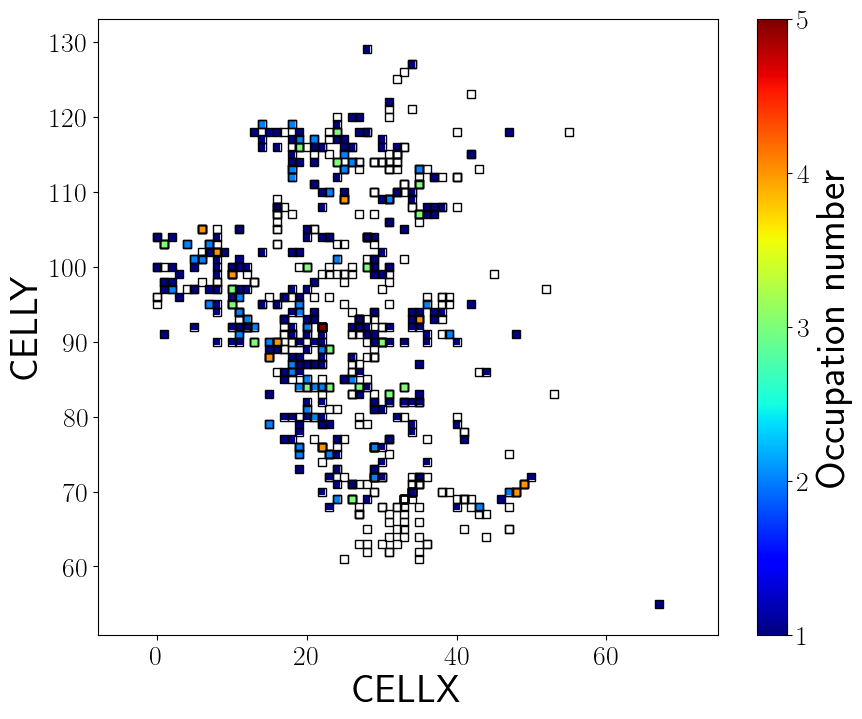}
        \includegraphics[scale=0.275]{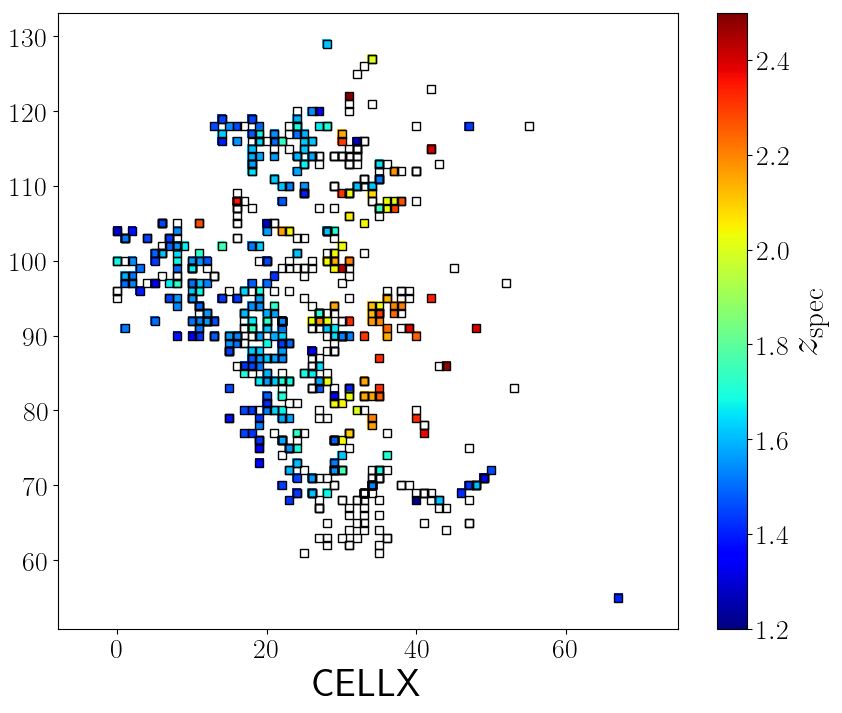}
        \includegraphics[scale=0.275]{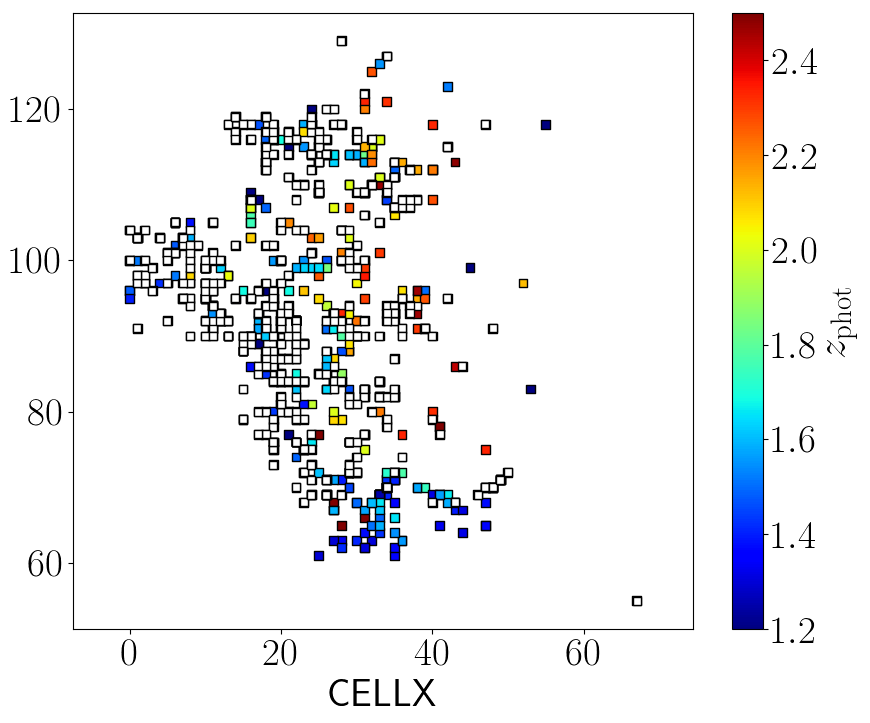}
    \caption{Representation of SOM cells targeted by the KMOS programme. {\it Left}: coloured cells are filled with high-quality spectroscopic redshift measurements in the three fields targeted by our survey, whereas empty cells are occupied by observed and not spectroscopically confirmed targets. The high-quality spectroscopically assigned cells are colour-coded according to the occupation level, meaning the number of validated galaxies occupying the same colour cell. {\it Middle}: the SOM cells filled with high-quality spectroscopic redshift measurements are colour-coded according to the assigned $z_{\rm spec}$. {\it Right}: the observed but still empty SOM cells are colour-coded according to the $z_{\rm phot}$ of the observed targets, whereas high-quality spectroscopic redshift measurements are coloured in white.}
    \label{fig:SOM_observed_validated_targets}
\end{figure*}

In the central panel of Fig. \ref{fig:SOM_observed_validated_targets}, validated cells are colour-coded according to the value of the assigned $z_{\rm spec}$. Cells populated with multiple observations have been assigned a median $z_{\rm spec}$ value.
This panel again highlights a prevalence of low-redshift targets as already discussed in Sect. \ref{Sec:results_I}, mainly concentrated at low values of the $X-$indices, and spread along the whole $Y-$index range.
In the right panel, we show the $z_{\rm phot}$ of the observed targets for which we could not measure $z_{\rm spec}$, and we mask the spectroscopically confirmed cells. 
The comparison between the $z_{\rm spec}$ and $z_{\rm phot}$ SOMs confirms that, despite the higher number of spectroscopically confirmed $H$-band targets, there is no systematic (photometric) redshift bias in the observed and non-validated targets: the SOM cells that were observed but could not be filled with a highly confident $z_{\rm spec}$ have values ranging from the lowest $H$-band to the highest redshifts reachable with the $K$-band setup.
However, if the lack of measurement is due to observational difficulties in the $K$-band and lower accuracy in the SED fitting $z_{\rm phot}$ determination used to select the observed targets, the cause of the concentration of lower redshift ($H$-band) galaxies present in the bottom region of the SOM (dark blue cells) must be investigated more thoroughly.

\begin{figure*}
    \centering
        \includegraphics[scale=0.275]{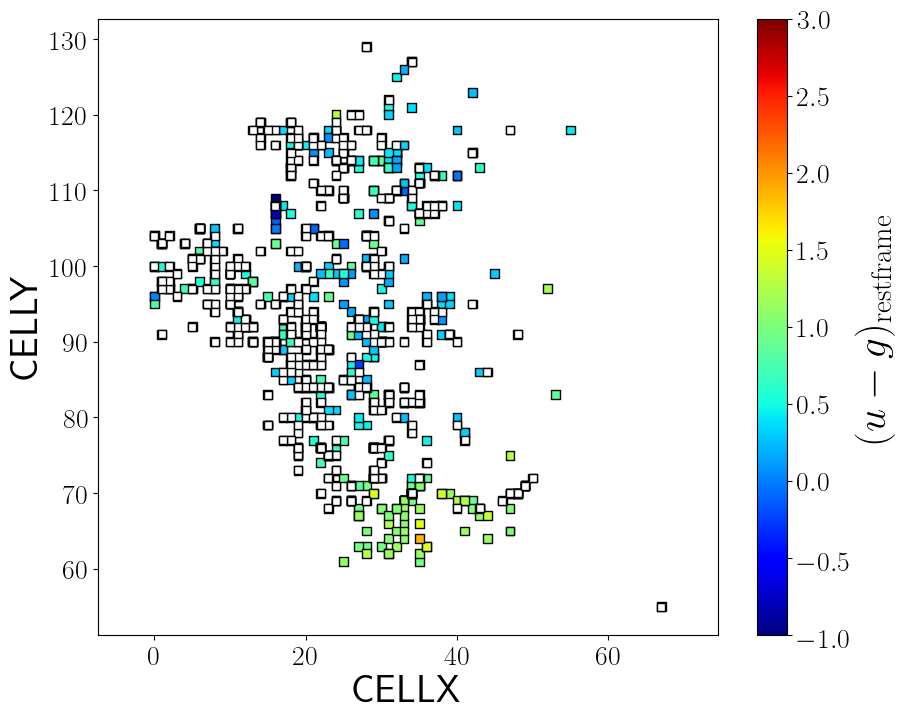}
        \includegraphics[scale=0.275]{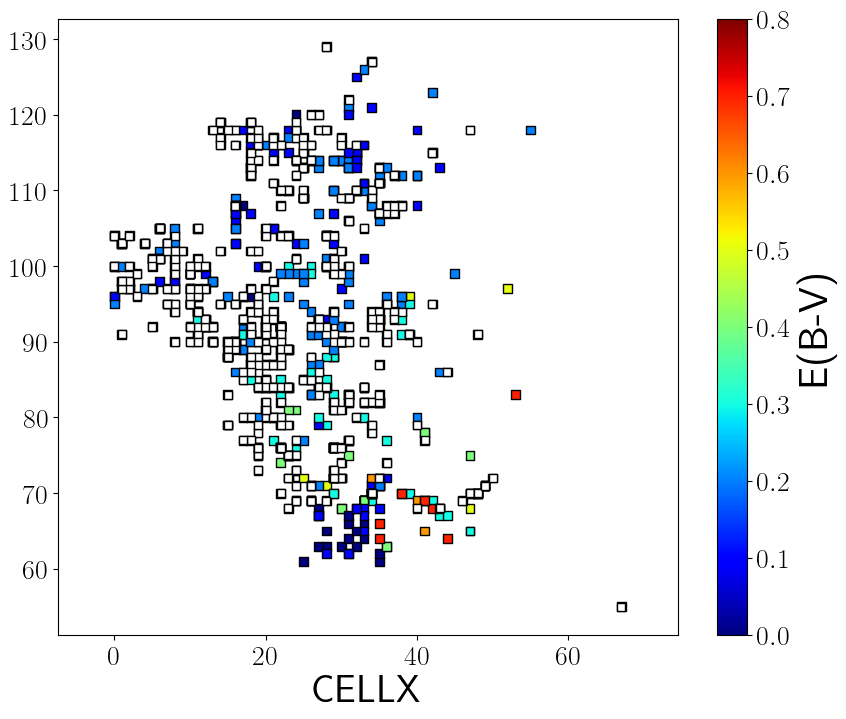}
        \includegraphics[scale=0.275]{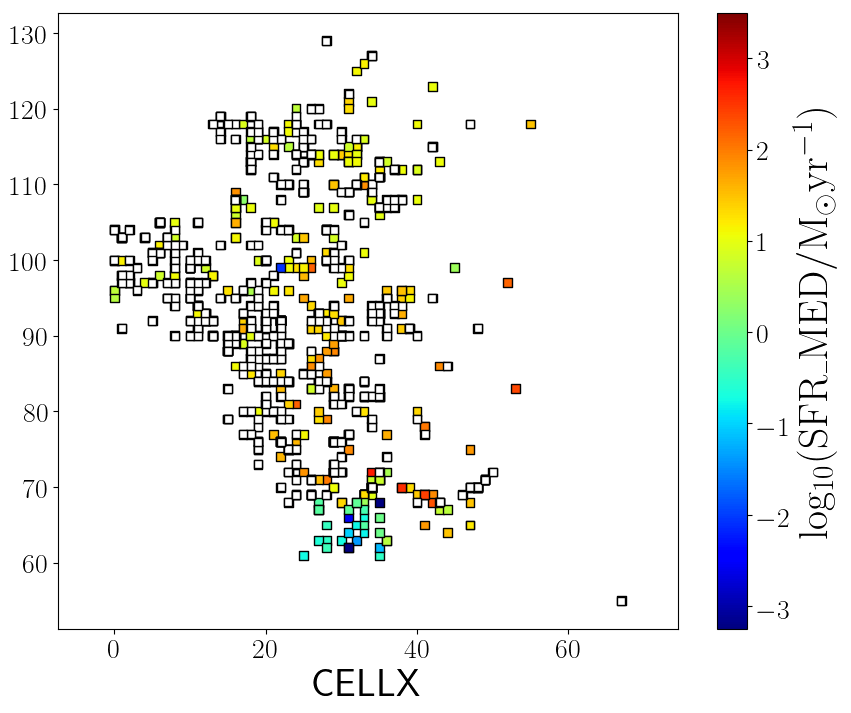}
    \caption{Representation of SOM cells targeted by the KMOS programme. The cells filled with high-quality spectroscopic redshift measurements are coloured in white. {\it Left}: cells are colour-coded according to the restframe $(u-g)$ colour. {\it Middle}: the cells are colour-coded according to the best fit $E(B-V)$ resulting from SED fitting analysis on the photometric sample. {\it Right}: the cells are colour-coded according to the best fit SFR resulting from SED fitting analysis on the photometric sample.}
    \label{fig:SOM_observed_properties}
\end{figure*}

We searched for the reason behind these spectroscopic failures in the colours and star formation properties of galaxies.
Figure \ref{fig:SOM_observed_properties} represents the rest-frame $(u-g)$ colour, the best fit $E(B-V),the $ and SED fitting SFR of the non-validated sample. Again, the cells containing more than one target have been assigned a median value.
The peculiarity of the bottom part of the SOM stands out: the galaxies populating these cells are, on average, redder and have lower star formation rates compared to the other empty cells. Moreover, as it noticeable from the $E(B-V)$ shown in the middle panel, they are not particularly dusty. Our observing strategy, and in particular the integration time, may require modifications for obtaining the necessary SNR required to measure emission-line redshifts.

\section{Results II: The physical properties of galaxies}
\label{Sec:results_II}

\subsection{SED fitting analysis}
\label{Sec:stellar_masses}

\begin{table}

\centering
\caption{Summary of the photometry used in each field. The complete filter set used in the COSMOS and SXDF data release is given in Table 1 of \cite{Laigle2016} and Table 1 of \cite{Mehta2018}. \label{Tab:filters}}
\begin{tabular}{cccc}
\hline
Field & Instrument/Telescope & Filter & Central \\
 & (Survey) & & $\lambda$ (\AA) \\
\hline
\hline
COSMOS & {\it GALEX} & NUV &  2313.9\\
 & MegaCam/CFHT & $u^\star$ & 3823.3\\
 & Suprime-Cam & $B$ & 4458.3\\
 & /Subaru & $V$ & 5477.8\\
 & & $r$ & 6288.7\\
 & & $i^+$ & 7683.9\\
 & & $z^{++}$ & 9105.7\\
 \hline
 & VIRCAM & $Y^{UD}$ & 10214.2\\
 & /VISTA & $J^{UD}$ & 12534.6\\
 & (Ultra VISTA-DR2) & $H^{UD}$ & 16453.4\\
 & & $K_S^{UD}$ & 21539.9\\
 \hline
 & IRAC/{\it Spitzer} & ch1 &  35634.3\\
 & (SPLASH) & ch2 & 45110.1\\
 & & ch3 & 57593.4\\
 & & ch4 &  79594.9\\
 \hline
 \hline
SXDF & MegaCam/CFHT & $u^\star$ & 3823.3\\
 & HSC & $g$& 4816\\
 & & $r$ & 6234\\
 & & $i$ & 7741\\
 & & $z$ & 8912\\
 & & $y$ & 9780\\
 \hline
 & VISTA & $Y$ & 10211\\
 & (VIDEO) & $J$ & 12541\\
 & & $H$ & 16464\\
 & & $K_S$ & 21488\\
 \hline
 & IRAC/{\it Spitzer} & ch1 &  35573\\
 & (SPLASH) & ch2 & 45049\\
 & & ch3 &  57386\\
 & & ch4 & 79274\\
 \hline
 \hline
VVDS & MegaCam/CFHT & $u$ & 3811\\
 & & $g$ & 4862\\
 & & $r$ & 6258\\
 & & $i$ & 7553\\
 & & $z$ & 8871\\
 \hline
 & VISTA & $Y$ & 10211\\
 & (VIDEO) & $J$ & 12541\\
 & & $H$ & 16464\\
 & & $K_S$ & 21488\\
\hline
\end{tabular}
\end{table}

The physical properties of galaxies were derived again for the spectroscopically confirmed targets, by taking advantage of the use of $z_{\rm spec}$ as a constraint to the fit.
We applied the SED fitting code \texttt{LePhare} to the spectrophometric catalogues obtained from merging the spectroscopic redshift measurement with the multi-band photometry collected from the parent surveys. A detailed list of the filters used in the three fields is reported in Table \ref{Tab:filters}, and the appropriate reference to the parent photometric catalogues is given in the table caption.
The code is provided with spectroscopic redshifts and total magnitudes as input, and we set the priors on fitting parameters and galaxy libraries (based on a collection of different star formation histories, SFHs) taking advantage of the knowledge of the average properties of our target galaxies: these are high-redshift, star-forming galaxies, with consistent H$\alpha$ emission. Out of the whole library of available models, we selected a number of exponentially declining SFHs ($\tau$ models), of delayed SFH and of constant SFR, with sub-solar ($Z=0.008$) and solar ($Z={\rm Z_\odot}=0.02$) metallicity.
We used a fine grid of $E(B-V)$ ranging from 0 to 0.7, and two different extinction laws \citep{Calzetti2000,Arnouts2013}, are also adopted. We obtain the stellar masses, absolute magnitudes, best fit $E(B-V)$ values, and other physical parameters such as the SFR as output. In the following, for stellar masses and SED fitting SFRs, we use the median values computed from the ML analysis of \texttt{LePhare}.

\begin{figure}
    \centering
    \includegraphics[scale=0.5]{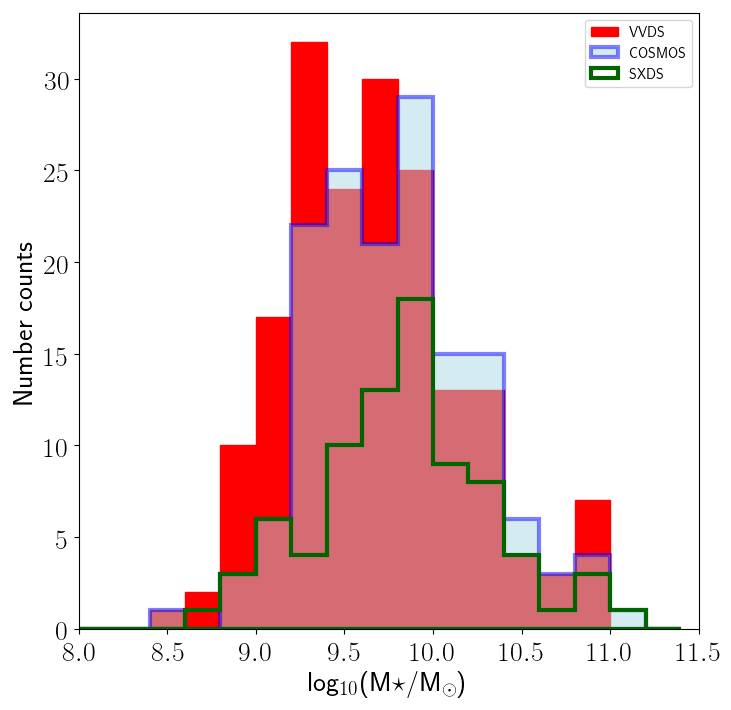}
    \caption{Histogram of stellar masses computed by \texttt{LePhare} on the spectrophotometric catalogues ($z_{\rm spec}$ sample) built in the three fields. The fields are shown with separate histograms as indicated by the legend.}
    \label{fig:stellar_mass_histo}
\end{figure}

The histogram of the resulting stellar masses from \texttt{LePhare} in the three fields is shown in Fig. \ref{fig:stellar_mass_histo}. The median stellar mass value in the total spectrophotometric sample of galaxies observed during the KMOS programme is $\logten (M_\star/\si{\solarmass})=9.69$, and the values in the three different fields are: $\logten (M_\star/\si{\solarmass})\rm_{COSMOS} = 9.73$, $\logten (M_\star/\si{\solarmass})\rm_{SXDF} = 9.84$, $\logten(M_\star/\si{\solarmass})\rm_{VVDS} = 9.62$.

Besides the primary goal of determining and calibrating $P(z\textbar \vec{C})$, the properties of the galaxies observed by the C3R2 survey is of unique importance and interest.
Building a sample of spectra spanning the whole redshift range up to $z\sim2.5$ and covering the whole galaxy colour space will shed light on controversial aspects of galaxy evolution studies and will help the acquisition of a general and complete picture of the galaxy zoology.
The KMOS C3R2 programme provides a number of physical properties of the spectroscopically confirmed galaxies, such as total H$\alpha$ fluxes and stellar masses.
In the following sections, we determine and discuss the physical properties of the spectroscopically confirmed galaxies in the COSMOS, VVDS, and SXDF fields, leaving aside the ECDFS field which contributes with only 12 galaxies to the release.

\subsection{H$\alpha$ fluxes}
\label{Sec:Halpha_flux}

The velocity and H$\alpha$ maps from \texttt{KUBEVIZ} allow the measurement of the total H$\alpha$ flux of the sources.
Starting from the centre coordinates, the final $z_{\rm spec}$ and the velocity map, we estimate the H$\alpha$ flux in a fixed circular aperture of 1\farcs.2 radius. This corresponds to about 10 kpc at redshifts $1.25 \lesssim z \lesssim 2.5$. \cite{vanderWel2014}, using 3D-HST (\HST) and CANDELS galaxies, as well as ACS/F814W (8073.43 \AA), WFC3/F125W (12501.04\,\AA), and WFC3/F160W (15418.27\,\AA) filters for measuring sizes, estimated the evolution of the effective radius ($R_{\rm e}$) of star-forming galaxies in various stellar mass and redshift bins. They estimated that massive star-forming galaxies ($M_\star \sim 10^{11}\, \si{\solarmass}$) have $R_{\rm e}\sim5\,{\rm kpc}$ in the redshift range probed by our KMOS survey. Thus, considering that the stellar mass distribution of our galaxy sample is below $10^{11}\,\si{\solarmass}$ (Fig. \ref{fig:stellar_mass_histo}), we considered an aperture from the galaxy centre that doubles the $R_{\rm e}$ estimated in \cite{vanderWel2014}. This way, we sample our sources up to the outskirts and obtain the total emission-line fluxes.

\begin{figure}
    \centering
    \includegraphics[scale=0.375]{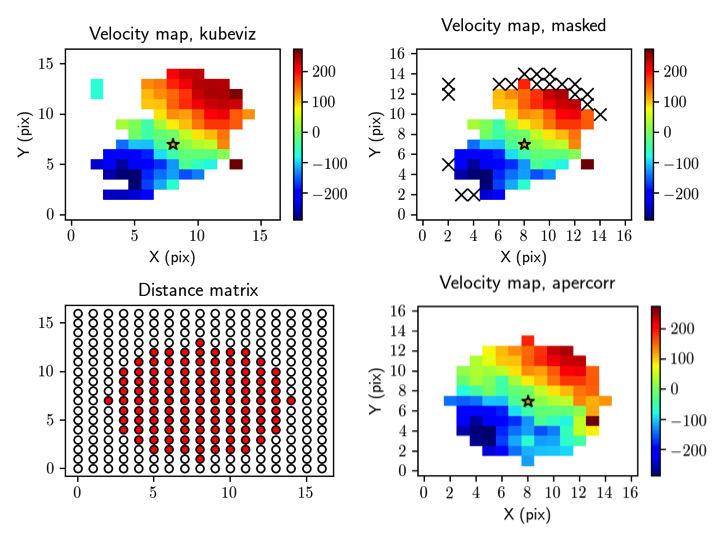}
    \caption{Summary of procedure followed to estimate the H$\alpha$ flux within the 1\farcs2 radius aperture, for a typical case of a galaxy with a rotation curve. The {\it top-left} panel shows the velocity map from \texttt{KUBEVIZ}. The star at the centre of the image reprensents the pixel position from which the aperture is estimated. The {\it bottom-left} panel shows the distance matrix that defines the six-pixel radius corresponding to the aperture. The {\it top-right} panel shows which spaxels from the original map are discarded because they fall outside the aperture. Finally, the {\it bottom-right} panel shows the corrected velocity field obtained following the procedure described in the main text for assigning a peculiar velocity to the spaxels flagged as bad in \texttt{KUBEVIZ}.}
    \label{fig:Halpha_apercorr}
\end{figure}

\begin{figure*}
    \centering
    \includegraphics[scale=0.45]{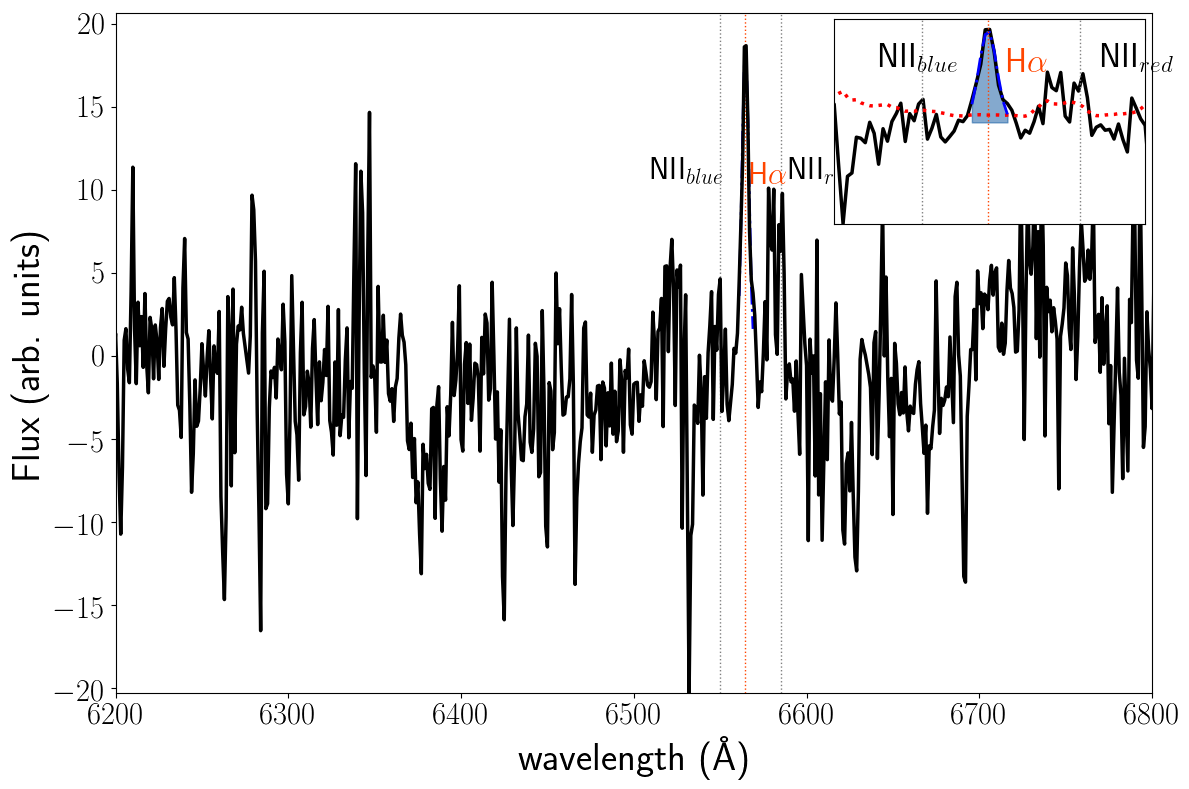}
    \caption{One-dimensional spectrum estimated by summing up all the spaxel spectra in the 1\farcs.2 radius aperture, corrected for their peculiar velocity according to the aperture-corrected velocity map described in the main text (Sect. \ref{Sec:Halpha_flux}). The same galaxy as the one shown in Fig. \ref{fig:Halpha_apercorr} is used. The main panel shows a wavelength cut of the whole 1D sum spectrum around the H$\alpha$ and N{\sc ii} lines, which are indicated with orange and black dashed lines, respectively. The inset panel is a zoom-in around the H$\alpha$ peak and shows the integral of the line that is estimated for measuring the total flux (light blue area) weighted by the noise (red dashed line), and it is also continuum corrected.}
    \label{fig:spec1d_apercorr}
\end{figure*}

A summary of the procedure followed for computing the H$\alpha$ aperture fluxes is shown, for a typical case of a galaxy with velocity field, in Fig. \ref{fig:Halpha_apercorr}.
We started from the velocity difference with respect to the galaxy centre estimated with \texttt{KUBEVIZ} and saved it as output in the velocity map (top-left panel of the figure). We also assigned a peculiar velocity to the spaxels entering the 1\farcs2 circular aperture (shown by means of a distance matrix in the top left panel of the figure) that were flagged as bad from the \texttt{KUBEVIZ} fit. This value is computed progressively as the mean of the peculiar velocities of the neighbouring spaxels, starting from the most populated (i.e. with the highest number of good fit neighbouring spaxels) regions in the map. This method, leading to the smooth velocity map in the aperture (shown in the bottom-right panel of the figure), assumes that the velocity curves we are considering are smooth (see \citealt{Wilman2020}), which is not a strong assumption for discy star-forming galaxies.
We then produced a total rest-frame 1D spectrum in the aperture by summing all the spaxels corrected for their relative velocity, as shown in Fig. \ref{fig:spec1d_apercorr} -- where the same galaxy of Fig. \ref{fig:Halpha_apercorr} is used. Furthermore, we estimated the integrated flux by performing a weighted Gaussian fit to the total rest-frame H$\alpha$ emission line, which was weighted for the noise spectrum. We subtracted the continuum contribution in two different ways. Firstly, we gave a rough estimate of the continuum of the spectrum as the median sigma clipped counts in two windows of 300 pixels in width blueward and redward of the emission line. Secondly, we considered the continuum on the H$\alpha$ emission as it was estimated by \texttt{KUBEVIZ} .
The method outlined above for measuring the H$\alpha$ emission-line flux does not take into account the H$\alpha$ stellar absorption, but this is small and can be neglected. Using synthetic spectra representative of our galaxy population (same redshift range, delayed SFHs in agreement with the \texttt{LePhare} best fit models), we estimate that the ratio between the equivalent width (EW) of the H$\alpha$ stellar absorption and the EW of the H$\alpha$ emission line (as measured from the KMOS data) is lower than 5\%.

\subsection{The SFR mass relation}

The H$\alpha$ flux is one of the primary SFR indicators, according to the well-known \cite{Kennicutt1998} relation, which sets a proportionality between H$\alpha$ flux and SFR, see Eq. \ref{eq:modified_kennicutt} below.
It is known that the extinction on the nebular emission is enhanced, on average, with respect to the extinction towards the stellar component, and several methods and calibrations have been performed to derive it. (1) Observed spectra covering a broad enough wavelength range allow the direct estimate of the absorption through the computation of observed emission-line ratios and their comparison to the theoretical value set by quantum physics, such as the ratio of the Balmer nebular emission lines H$\alpha$/H$\beta$. (2) A number of relations linking the absorption in the continuum to that in the emission lines \citep{Calzetti2000,Wuyts2013} have been studied at various redshift and in different wavelength regimes over the last few years (3) Finally, the Kennicutt SFR--H$\alpha$ relation has also been calibrated by means of multiple SFR indicators to derive the best fit nebular extinction value \emph{a posteriori}, such as the work performed in \cite{Kashino2019}.

\begin{figure*}
    \centering
    \includegraphics[scale=0.35]{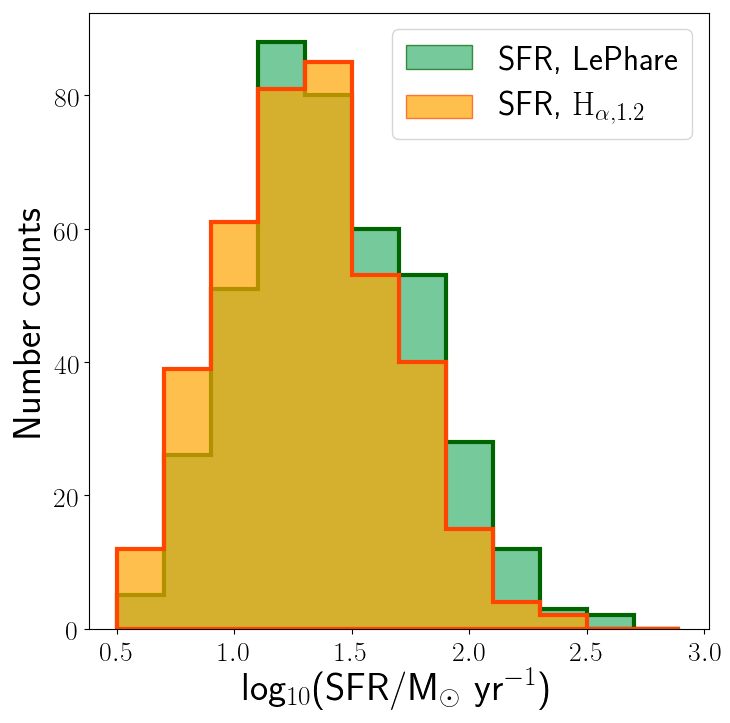}
    \includegraphics[scale=0.35]{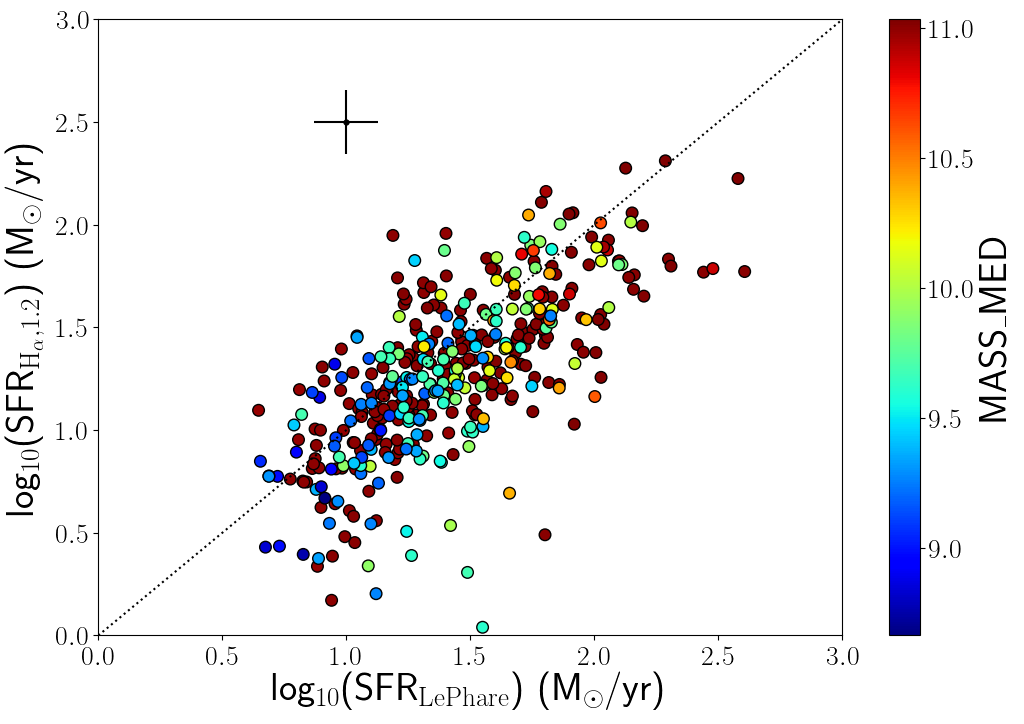}
    \caption{{\it Left}: histogram of SFR derived from aperture H$\alpha$ fluxes, and that estimated from \texttt{LePhare} SED fitting. {\it Right}: comparison between the H$\alpha$ and SED-fitting SFRs, colour-coded by galaxy stellar mass. The black dashed line is the one-to-one correlation. The plot also shows the typical error on the SFR from \texttt{LePhare} (horizontal black error bar, calculated using the SFR\_INF and SFR\_SUP released in the catalogue) and on the H$\alpha$ SFR (considering a typical uncertainty of 10\% on the flux measurement, see \citealt{Wisnioski2019}).}
    \label{fig:SFR_aperture_LePhare}
\end{figure*}

Considering the items above, the \cite{Kennicutt1998} equation, for a \cite{Chabrier2003} IMF, becomes:
\begin{equation}
F_{{\rm H}\alpha}[{\rm erg \, cm^{-1} \, s^{-1}}] = \frac{{\rm SFR} \,[\si{\solarmass} \, {\rm yr^{-1}}]}{4.6 \times 10^{-42}} \cdot \frac{1}{4 \pi d_{\rm L}^2} \cdot 10^{-0.4A_{{\rm H}\alpha}} , 
\label{eq:modified_kennicutt}
\end{equation}
\noindent where $d_{\rm L}$ is the luminosity distance, and
\begin{equation}
A_{{\rm H}\alpha} = K_{{\rm H}\alpha} \frac{E(B-V)}{{\rm f_{neb}}}.
\label{eq:Halpha_absorption}
\end{equation}
$K_{{\rm H}\alpha}=2.54$ is the wavelength dependence of extinction according to \cite{Cardelli1989}, $E(B-V)$ is the reddening resulting from \texttt{LePhare,} and ${\rm f_{neb}}=0.53\pm 0.01$ is the enhancement of extinction towards nebular lines calibrated in \cite{Kashino2019}.
The error associated with each object is 0.15\,dex, and it is added in quadrature to the typical error associated to the flux measurement (vertical error bar in Figure \ref{fig:SFR_aperture_LePhare}).
We derived SFR using Eq. (\ref{eq:modified_kennicutt}) with the H$\alpha$ aperture fluxes (Sect. \ref{Sec:Halpha_flux}) and the luminosity distance based on the spectroscopic redshift measurements.

Figure \ref{fig:SFR_aperture_LePhare} shows the resulting H$\alpha$-based SFRs compared with those estimated from SED fitting with \texttt{LePhare}. Both distributions peak at $\logten({\rm SFR}/\si{\solarmass} {\rm yr^{-1}})\sim$\,1.0 -- 1.5, but SED-fitting SFRs are systematically higher than those from aperture H$\alpha$ fluxes (of the order of 0.05-0.1\,dex in each of the three fields).
We point out that the SFRs derived with LePhare are instantaneous, in agreement with the definition of a H$\alpha$-based SFR. However, differences may arise from (1) the necessary approximations adopted in the SED-fitting procedure in order to derive SFRs as well as other physical parameters (e.g. the number of input SED, the limited number of ages in the grid); (2) the uncertainties in the extinction values derived through the SED fitting (see \cite{Laigle2019} for details); and (3) the uncertainties in the relation between continuum and line absorption that we had to adopt to derive the SFR from H$\alpha$ fluxes.
Furthermore, in light of the considerations previously performed on the sizes of our galaxy sample, this systematic shift is not likely to be attributable to the different area considered in the photometry with respect to the aperture considered for computing the total H$\alpha$ flux. Indeed, as is noticeable from the stellar mass distribution, these galaxies are less massive than those considered as a reference for choosing the appropriate flux aperture. Moreover, SFRs derived from SED fitting are compatible with the scatter of the plot around the 1:1 line (approximately 0.5\,dex).

\begin{figure*}
    \centering
    \includegraphics[scale=0.65]{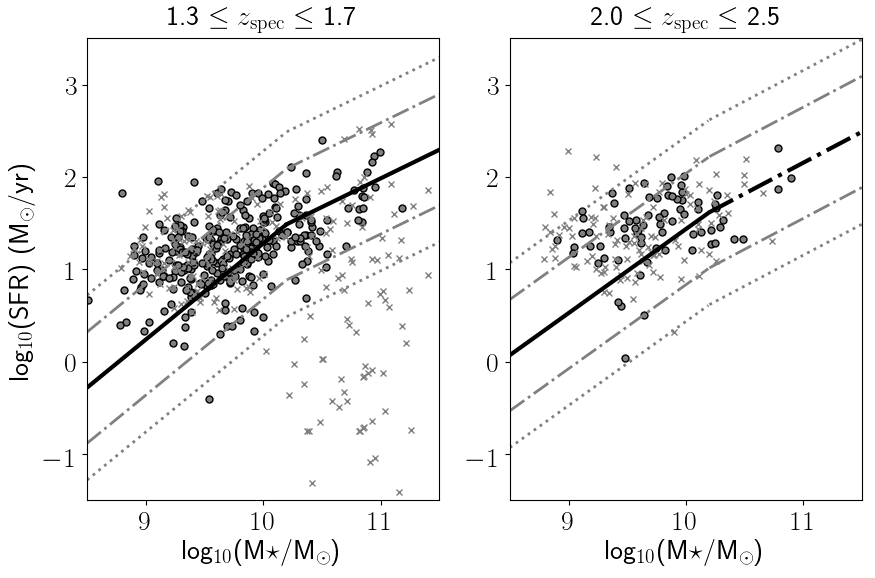}
    \caption{(H$_\alpha-$ based) SFR (grey circles) and (SED fitting based) SFR (grey crosses) $vs$ stellar mass. The left panel shows the lower redshift targets observed in $H$-band in the three surveys considered in the scientific analysis prensented here, and the {\it right} panel represents the same for higher redshift $K$-band targets. The black solid lines are the best fit to the star-forming main sequence (MS) in the same redshift range from \cite{Whitaker2014}; the dashed and dotted lines show 4$\times$ and 10$\times$ above and below the MS and bracket the distribution of the data points of the 3D-HST galaxies (see Fig. 7 in \citealt{Wisnioski2019}).}
    \label{fig:SFR_Mass_HK}
\end{figure*}

The distribution of the derived SFR and stellar masses in the SFR mass plane is shown in Fig. \ref{fig:SFR_Mass_HK}.
The star-forming main sequence (MS, black dashed line) parametrisation adopted is a  broken power law defined in the stellar mass range $9.2\leq \logten(M_\star/\si{\solarmass}) \leq 11.2$ using UV and infrared SFRs from 3D-HST data at $0.5 \leq z \leq 2.5$ in all CANDELS fields \citep{Whitaker2014}. In the $H$-band, the $\rm SFR  mass$ relation is lower than that at higher redshift ($K$-band). In particular, the distribution of both the KMOS $H-$ and $K$-band sources is systematically higher than the star-forming main sequence.
As already discussed in the SR analysis (Fig. \ref{fig:histo_observed_validated_targets}, bottom-right panel), this trend indicates that due to the low stellar mass of the galaxies observed, the SR is biased towards highly star-forming galaxies above the MS.

In the figure, we also included the SED-fitting-based SFR of non-validated galaxies (grey crosses). As is noticeable, at $1.3 \leq z \leq 1.7$ ($H$-band), the population of low star-forming galaxies previously identified in Sect. \ref{subsec:spec_failures} emerges; the distribution of grey crosses at $2.0 \leq z \leq 2.5$ ($K$-band) is not remarkably different from that of spectroscopically confirmed targets (grey circles), further confirming that spectroscopic failures in this regime are more likely due to higher uncertainties in $z_{\rm phot}$.

The KMOS\MVAt C3R2 stellar mass distribution peaks at $\logten(M_\star/\si{\solarmass})\sim 9.5$, which corresponds to the lower edge of the stellar mass distribution of the KMOS 3D galaxies \citep{Wisnioski2019}.
The integration between the two samples lays the groundwork for building a high-redshift SFR mass relation that is able to probe a wider stellar mass range, with the ultimate goal of determining the characteristic mass above which a flattening of the MS relation is expected to occur (\citealt{Elbaz2007} at $z\sim 1$; \citealt{Daddi2007} at $z\sim 2$).

\section{Catalogue release}
\label{sec:catalogue}
Following the methodology outlined above, we built a table containing the redshift assigned in each of the observed pointings, together with some relevant information regarding the observed targets. The released catalogue collects all high-quality ($Q \geq 3$) redshift measurements.
Below, we describe the columns of the catalogue. The properties of a sub-sample of galaxies are given in Table \ref{tab:spec_z}, while the total sample can be found at CDS.

The columns indicate the following parameters:
\begin{enumerate}
\item OBJ\_ID: identification number for galaxies
\item RA: right ascension (deg)
\item Dec: declination (deg)
\item Pointing: name of the KMOS OB in which the galaxy has been observed (see Table \ref{Tab:observed_pointings})
\item Z\_SPEC: redshift assigned and validated as described in Sect. \ref{Sec:redshift_assignment}
\item Q\_flag: quality flag of the redshift measurement, assigned according to the criteria described in Sect. \ref{Sec:redshift_assignment}
\item PHOTO-Z: photometric redshift from the galaxy parent survey (details are given in Sect. \ref{Sec:photometric_catalogues})
\item Priority (M17): observational priority of the target, according to the scheme described in M17
\item EBV\_BEST: $E(B-V)$ computed with \texttt{LePhare}
\item MASS\_INF: sixteenth percentile of the galaxy stellar mass from the maximum likelihood (ML) analysis of \texttt{LePhare}
\item MASS\_MED: median value of the galaxy stellar mass from the ML analysis of \texttt{LePhare}
\item MASS\_SUP: eighty-fourth percentile of the galaxy stellar mass from the ML analysis of \texttt{LePhare}
\item SFR\_INF: sixteenth percentile of the SFR from the maximum likelihood (ML) analysis of \texttt{LePhare}
\item SFR\_MED: median value of the SFR from the ML analysis of \texttt{LePhare}
\item SFR\_SUP: eighty-fourth percentile of the SFR from the ML analysis of \texttt{LePhare}
\item F$_{{\rm H}\alpha,1.2}$: H$\alpha$ flux computed within an aperture of 1\farcs2 radius (see Sect. \ref{Sec:Halpha_flux})
\end{enumerate}

\begin{table*}
\begin{center}
\caption{Sub-sample of ten galaxies in the catalogue with their properties. The full table can be found at CDS. The explanation of the different columns is given in Sect. \ref{sec:catalogue}. The column `ID' is repeated at the beginning of each part of the table for the sake of clarity. We estimated that, due to the uncertainties in the spectrophotometric calibrations, the precision on the H$\alpha$ flux measurement is not better than 10\%.
\label{tab:spec_z}}
\begin{tiny}

\begin{tabular}{lrrrrr}
\hline
  \multicolumn{1}{|c|}{OBJ\_ID} &
  \multicolumn{1}{c|}{RA} &
  \multicolumn{1}{c|}{Dec} &
  \multicolumn{1}{c|}{Pointing} &
  \multicolumn{1}{c|}{Z\_SPEC} &
  \multicolumn{1}{c|}{Q\_flag} \\
  \hline
  373952 & 150.36320 & 2.46340 & P100\_COSMOS\_HaHP1 & 1.7195 & 4.0\\
  399202 & 150.37578 & 2.51607 & P100\_COSMOS\_HaHP1 & 1.5130 & 4.0\\
  399761 & 150.34360 & 2.51690 & P100\_COSMOS\_HaHP1 & 1.3991 & 4.0\\
  388984 & 34.74180 & $-4.86346$ & P100\_SXDF\_HaKP2 & 2.3486 & 3.0\\
  105609 & 34.59267 & $-5.35292$ & P100\_SXDF\_haHP1\_v2 & 1.6199 & 3.0\\
  111251 & 34.61345 & $-5.34167$ & P100\_SXDF\_haHP1\_v2 & 1.5970 & 4.0\\
  122473 & 34.61895 & $-5.31527$ & P100\_SXDF\_haHP1\_v2 & 1.6256 & 3.0\\
  274911 & 36.720165 & $-4.46552$ & P100\_VVDS\_HaHP2 & 1.6092 & 4.0\\
  394673 & 36.31682 & $-4.244806$ & P102\_VVDS\_HaHP2 & 1.5689 & 4.0\\
  247070 & 36.87232 & $-4.51824$ & P101\_VVDS\_HaHP2 & 1.5130 & 4.0\\
  390870 & 36.33725 & $-4.25240$ & P99\_VVDS\_HaHP2\_v2 & 1.4341 & 3.5\\
  \hline
\vspace{2.5mm}
\end{tabular}

\begin{tabular}{lrrr}
\hline
  \multicolumn{1}{|c|}{OBJ\_ID} &
  \multicolumn{1}{c|}{PHOTO-Z} &
  \multicolumn{1}{c|}{priority (M17)} &
  \multicolumn{1}{c|}{EBV\_BEST} \\
  \hline
  373952 & 1.5269 & 500 & 0.1\\
  377914 & 1.6315 & 250 & 0.1\\
  399202 & 1.4793 & 500 & 0.3\\
  399761 & 1.3806 & 400 & 0.3\\
  400978 & 1.5295 & 1000 & 0.3\\
  405462 & 1.5557 & 200 & 0.3\\
  405597 & 1.4334 & 250 & 0.3\\
  405666 & 1.4477 & 200 & 0.2\\
  405763 & 1.4553 & 1000 & 0.5\\
  \hline
  \vspace{2.5mm}
\end{tabular}

\begin{tabular}{lrrrrrrr}
\hline
  \multicolumn{1}{|c|}{OBJ\_ID} &
  \multicolumn{1}{c|}{MASS\_INF} &
  \multicolumn{1}{c|}{MASS\_MED} &
  \multicolumn{1}{c|}{MASS\_SUP} &
  \multicolumn{1}{c|}{SFR\_INF} &
  \multicolumn{1}{c|}{SFR\_MED} &
  \multicolumn{1}{c|}{SFR\_SUP} &
  \multicolumn{1}{c|}{$\rm F_{{\rm H}\alpha,1.2}$} \\
  \multicolumn{1}{|c|}{} &
  \multicolumn{1}{c|}{$\logten(M_\star/\si{\solarmass})$} &
  \multicolumn{1}{c|}{$\logten(M_\star/\si{\solarmass})$} &
  \multicolumn{1}{c|}{$\logten(M_\star/\si{\solarmass})$} &
  \multicolumn{1}{c|}{$\logten(\si{\solarmass}\, yr^{-1})$} &
  \multicolumn{1}{c|}{$\logten(\si{\solarmass}\, yr^{-1})$} &
  \multicolumn{1}{c|}{$\logten(\si{\solarmass}\, yr^{-1})$} &
  \multicolumn{1}{c|}{$\rm 10^{-17}\,erg\, cm^{-2}\, s^{-1}$} \\
\hline
  373952 & 8.93 & 9.15 & 9.25 & 0.89 & 1.04 & 1.43 & 4.44\\
  377914 & 9.10 & 9.35 & 9.49 & 0.83 & 0.98 & 1.33 & 4.45\\
  399202 & 10.01 & 10.11 & 10.18 & 0.95 & 1.15 & 1.38 & 7.87\\
  399761 & 9.80 & 10.06 & 10.14 & 1.11 & 1.24 & 1.57 & 6.98\\
  400978 & 9.23 & 9.53 & 9.68 & 1.06 & 1.36 & 1.50 & 5.13\\
  405462 & 9.95 & 10.00 & 10.05 & 1.33 & 1.44 & 1.55 & 8.10\\
  405597 & 10.37 & 10.41 & 10.44 & 1.72 & 1.80 & 1.88 & 12.91\\
  405666 & 9.66 & 9.73 & 9.80 & 1.01 & 1.11 & 1.19 & 10.22\\
  405763 & 10.18 & 10.36 & 10.46 & 1.47 & 1.68 & 1.93 & 3.24\\
  \hline
  \vspace{2.5mm}
\end{tabular}

 \end{tiny}
 
 \end{center}
\end{table*}

\section{Conclusions}
\label{Sec:conclusions}

In this work, we present the first results of a 200\,h ESO Large Programme (199.A-0732; PI F.J. Castander) consisting of VLT spectroscopic observations, as part of the C3R2 survey.
The main goal of C3R2 is to acquire accurate spectroscopic redshifts across the relevant galaxy colour space in order to accurately determine the colour-redshift relation for the {\it Euclid} weak lensing cosmological survey.
As a contribution to this challenging goal, we release a spectrophotometric catalogue of high-redshift star-forming galaxies observed for 88\,h with the near-infrared KMOS spectrograph.
A total of 424 high-quality spectroscopic redshifts have been determined over five semesters in four extragalactic fields (COSMOS, SXDF, ECDFS, and VVDS-02h), mainly measured as single emission-line redshifts ($\rm Q \leq 3.5$) in two near-infrared filters: the $H$ (1.456 -- 1.846\, \micron) filter allows us to detect H$\alpha$ ($\lambda=6564.61$ \AA) at $1.3 \leq z \leq 1.7,$ and the $K$ (1.934 -- 2.460\,\micron) filter allows us to detect H$\alpha$ at $2.0 \leq z \leq 2.5$.
Of the 424 high-quality spectroscopic redshifts assigned, 255 (60\%) are based on single emission-line identification (or multiple emission lines with an unsatisfactory SNR), and the remaining 40\% were computed using multiple lines.
The main results can be divided in two categories,which we summarise below.
\subsubsection*{\it The spectroscopic SR}
A total number of 150 new redshifts were measured to galaxies belonging to the COSMOS field, 81 redshifts to galaxies belonging to the SXDF field, and 181 to galaxies in the VVDS-02h field, with an overall SR of 60.7\% for $H$-band observations and 32.8\% for $K$-band observations. 
We divided our target galaxies into two priority classes (P1 and P2). We were able to fill the 57\% of the observed P1 empty cells of the galaxy colour SOM, and 70\% of the observed P2 empty cells.
In Fig. \ref{fig:SR_cells}, we notice that less than 4\% of P1 cells and about 50\% of P2 cells in the near-infrared domain remain unexplored. However, 18 out of the total 269 cells we filled presented some problems in terms of $z_{phot}$ distribution, so they need to be investigated further, and possibly excluded from the {\it Euclid} calibration sample.
Considering our spectroscopic failures, we found that they mainly include (1) $K$-band targets whose SR is lower due to observational difficulties and lower accuracy of the $z_{\rm phot}$ estimate used at the sample selection stage, and (2) $H$-band galaxies with redder colours and lower SFR, which are more difficult to detect with the 1\,h integration time adopted by our observations. 

A follow-up near-infrared observing programme is ongoing with the Large Bincocular Telescope (LBT), making use of the two multi-object spectrographs LUCI1 and LUCI2. Our observing strategy is to simultaneously observe the same pointing using $H-$ and $K-$ band masks with LUCI1 and LUCI2, maintaining the same integration time of KMOS observations (1\,h). This allows us to observe many galaxies simultaneously in both filters, and helps us understand how much of the non-detection can be assessed with a broader wavelength range in the spectrum (e.g. in case of the more insecure photo-z estimates in $K$ band targets).

\subsubsection*{ \it The physical properties of the released galaxies}

We measured the physical properties of the spectroscopically confirmed galaxies using their KMOS resolved spectra and their optical and near-infrared photometry from public data release catalogues in the three fields.
We measured total H$\alpha$ fluxes in 1\farcs2 radius apertures from the total 1D spectrum obtained after correcting each spaxel for its peculiar velocity, and we computed other physical parameters such as stellar masses, absolute magnitudes, and extinction from SED fitting with fixed spectroscopic redshift.
The stellar mass distribution of our sample peaks at $\logten(M_\star/\si{\solarmass})=9.69$ and is similar within the error bars across the three fields.
We finally derived SFRs from the aperture H$\alpha$ flux following the \cite{Kennicutt1998} prescription, taking into account enhanced extinction towards nebular lines in the star-forming regions according to \citealt{Kashino2019}.
We studied the distribution of our galaxies in the SFR mass plane and compared our data points with the best fit high-redshift main sequence from \cite{Whitaker2014}. Galaxies observed during our KMOS programme are located, on average, at higher SFRs with respect to the average population of similar stellar masses. This result is due, especially at low stellar masses, to the limitations imposed by our observing strategy, of which the primary goal was to maximise the number of spectroscopic redshifts measured.
The peculiarity of our galaxy sample with respect to the literature, and in particular with respect to the KMOS-3D survey, is the stellar mass regime exploited. Our galaxies are, on average, less massive than those observed in KMOS-3D, and could be used as a starting point for future studies aiming to probe the lower stellar mass regime of the high-redshift SFR mass relation.

\begin{acknowledgements}
The Euclid Consortium acknowledges the European Space Agency and the support of a number of agencies and institutes that have supported the development of {\it Euclid}. A detailed complete list is available on the {\it Euclid} web site (\url{http://www.euclid-ec.org}). In particular the Academy of Finland, the Agenzia Spaziale Italiana, the Belgian Science Policy, the Canadian Euclid Consortium, the Centre National d'Etudes Spatiales, the Deutsches
Zentrum f\"{u}r Luft- und Raumfahrt, the Danish Space Research Institute, the Funda\c{c}\~{a}o para a Ci\^{e}nca e a Tecnologia, the Ministerio de Economia y Competitividad, the National Aeronautics and Space Administration, the Netherlandse Onderzoekschool Voor Astronomie, the Norvegian Space Center, the Romanian Space Agency, the State Secretariat for Education, Research and Innovation (SERI) at the Swiss Space Office (SSO), and the United Kingdom Space Agency.
Based on observations collected at the European Southern Observatory under ESO programme 199.A-0732 (B,D,F,H).
VG, RS, AG and RB acknowledge support by the Deutsches
Zentrum f\''{u}r Luft- und Raumfahrt (DLR) grant 50 QE 1101.
FJC acknowledges support from the Spanish Ministry of Science, 
Innovation and Universities through grant ESP2017-89838-C3-1-R,
and the H2020 programme of the European Commission through grant 776247.
AG acknowledges a Sinergia grant from the Swiss National Science Foundation.
SA thank the support PRIN MIUR 2015 ``Cosmology and Fundamental Physics: Illuminating the Dark Universe
with Euclid".
\end{acknowledgements}

\bibliographystyle{aa}
\bibliography{bibliography.bib}

\begin{thebibliography}{47}
\expandafter\ifx\csname natexlab\endcsname\relax\def\natexlab#1{#1}\fi

\bibitem[{{Aihara} {et~al.}(2017){Aihara}, {Armstrong}, {Bickerton}, {Bosch},
  {Coupon}, {Furusawa}, {Hayashi}, {Ikeda}, {Kamata}, {Karoji}, {Kawanomoto},
  {Koike}, {Komiyama}, {Lupton}, {Mineo}, {Miyatake}, {Miyazaki}, {Morokuma},
  {Obuchi}, {Oishi}, {Okura}, {Price}, {Takata}, {Tanaka}, {Tanaka}, {Tanaka},
  {Uchida}, {Uraguchi}, {Utsumi}, {Wang}, {Yamada}, {Yamanoi}, {Yasuda},
  {Arimoto}, {Chiba}, {Finet}, {Fujimori}, {Fujimoto}, {Furusawa}, {Goto},
  {Goulding}, {Gunn}, {Harikane}, {Hattori}, {Hayashi}, {Helminiak}, {Higuchi},
  {Hikage}, {Ho}, {Hsieh}, {Huang}, {Huang}, {Imanishi}, {Iwata}, {Jaelani},
  {Jian}, {Kashikawa}, {Katayama}, {Kojima}, {Konno}, {Koshida}, {Kusakabe},
  {Leauthaud}, {Lee}, {Lin}, {Lin}, {Mandelbaum}, {Matsuoka}, {Medezinski},
  {Miyama}, {Momose}, {More}, {More}, {Mukae}, {Murata}, {Murayama}, {Nagao},
  {Nakata}, {Niikura}, {Nishizawa}, {Oguri}, {Okabe}, {Ono}, {Onodera},
  {Onoue}, {Ouchi}, {Pyo}, {Shibuya}, {Shimasaku}, {Simet}, {Speagle},
  {Spergel}, {Strauss}, {Sugahara}, {Sugiyama}, {Suto}, {Suzuki}, {Tait},
  {Takada}, {Terai}, {Toba}, {Turner}, {Uchiyama}, {Umetsu}, {Urata}, {Usuda},
  {Yeh}, \& {Yuma}}]{Aihara2017}
{Aihara}, H., {Armstrong}, R., {Bickerton}, S., {et~al.} 2017, ArXiv e-prints,
  arXiv:1702.08449

\bibitem[{{Amara} \& {R{\'e}fr{\'e}gier}(2007)}]{Amara2007}
{Amara}, A. \& {R{\'e}fr{\'e}gier}, A. 2007, \mnras, 381, 1018

\bibitem[{{Arnouts} {et~al.}(1999){Arnouts}, {Cristiani}, {Moscardini},
  {Matarrese}, {Lucchin}, {Fontana}, \& {Giallongo}}]{Arnouts1999}
{Arnouts}, S., {Cristiani}, S., {Moscardini}, L., {et~al.} 1999, \mnras, 310,
  540

\bibitem[{{Arnouts} {et~al.}(2013){Arnouts}, {Le Floc'h}, {Chevallard},
  {Johnson}, {Ilbert}, {Treyer}, {Aussel}, {Capak}, {Sanders}, {Scoville},
  {McCracken}, {Milliard}, {Pozzetti}, \& {Salvato}}]{Arnouts2013}
{Arnouts}, S., {Le Floc'h}, E., {Chevallard}, J., {et~al.} 2013, \aap, 558, A67

\bibitem[{{Calzetti} {et~al.}(2000){Calzetti}, {Armus}, {Bohlin}, {Kinney},
  {Koornneef}, \& {Storchi-Bergmann}}]{Calzetti2000}
{Calzetti}, D., {Armus}, L., {Bohlin}, R.~C., {et~al.} 2000, \apj, 533, 682

\bibitem[{{Capak} {et~al.}(2007){Capak}, {Aussel}, {Ajiki}, {McCracken},
  {Mobasher}, {Scoville}, {Shopbell}, {Taniguchi}, {Thompson}, {Tribiano},
  {Sasaki}, {Blain}, {Brusa}, {Carilli}, {Comastri}, {Carollo}, {Cassata},
  {Colbert}, {Ellis}, {Elvis}, {Giavalisco}, {Green}, {Guzzo}, {Hasinger},
  {Ilbert}, {Impey}, {Jahnke}, {Kartaltepe}, {Kneib}, {Koda}, {Koekemoer},
  {Komiyama}, {Leauthaud}, {Le Fevre}, {Lilly}, {Liu}, {Massey}, {Miyazaki},
  {Murayama}, {Nagao}, {Peacock}, {Pickles}, {Porciani}, {Renzini}, {Rhodes},
  {Rich}, {Salvato}, {Sanders}, {Scarlata}, {Schiminovich}, {Schinnerer},
  {Scodeggio}, {Sheth}, {Shioya}, {Tasca}, {Taylor}, {Yan}, \&
  {Zamorani}}]{Capak2007}
{Capak}, P., {Aussel}, H., {Ajiki}, M., {et~al.} 2007, \apjs, 172, 99

\bibitem[{{Cardelli} {et~al.}(1989){Cardelli}, {Clayton}, \&
  {Mathis}}]{Cardelli1989}
{Cardelli}, J.~A., {Clayton}, G.~C., \& {Mathis}, J.~S. 1989, \apj, 345, 245

\bibitem[{{Chabrier}(2003)}]{Chabrier2003}
{Chabrier}, G. 2003, \pasp, 115, 763

\bibitem[{{Daddi} {et~al.}(2007){Daddi}, {Dickinson}, {Morrison}, {Chary},
  {Cimatti}, {Elbaz}, {Frayer}, {Renzini}, {Pope}, {Alexander}, {Bauer},
  {Giavalisco}, {Huynh}, {Kurk}, \& {Mignoli}}]{Daddi2007}
{Daddi}, E., {Dickinson}, M., {Morrison}, G., {et~al.} 2007, \apj, 670, 156

\bibitem[{{Dahlen} {et~al.}(2013){Dahlen}, {Mobasher}, {Faber}, {Ferguson},
  {Barro}, {Finkelstein}, {Finlator}, {Fontana}, {Gruetzbauch}, {Johnson},
  {Pforr}, {Salvato}, {Wiklind}, {Wuyts}, {Acquaviva}, {Dickinson}, {Guo},
  {Huang}, {Huang}, {Newman}, {Bell}, {Conselice}, {Galametz}, {Gawiser},
  {Giavalisco}, {Grogin}, {Hathi}, {Kocevski}, {Koekemoer}, {Koo}, {Lee},
  {McGrath}, {Papovich}, {Peth}, {Ryan}, {Somerville}, {Weiner}, \&
  {Wilson}}]{Dahlen2013}
{Dahlen}, T., {Mobasher}, B., {Faber}, S.~M., {et~al.} 2013, \apj, 775, 93

\bibitem[{{Dark Energy Survey Collaboration} {et~al.}(2016){Dark Energy Survey
  Collaboration}, {Abbott}, {Abdalla}, {Aleksi{\'c}}, {Allam}, {Amara},
  {Bacon}, {Balbinot}, {Banerji}, {Bechtol}, {Benoit-L{\'e}vy}, {Bernstein},
  {Bertin}, {Blazek}, {Bonnett}, {Bridle}, {Brooks}, {Brunner}, {Buckley-Geer},
  {Burke}, {Caminha}, {Capozzi}, {Carlsen}, {Carnero-Rosell}, {Carollo},
  {Carrasco-Kind}, {Carretero}, {Castander}, {Clerkin}, {Collett}, {Conselice},
  {Crocce}, {Cunha}, {D'Andrea}, {da Costa}, {Davis}, {Desai}, {Diehl},
  {Dietrich}, {Dodelson}, {Doel}, {Drlica-Wagner}, {Estrada}, {Etherington},
  {Evrard}, {Fabbri}, {Finley}, {Flaugher}, {Foley}, {Fosalba}, {Frieman},
  {Garc{\'\i}a-Bellido}, {Gaztanaga}, {Gerdes}, {Giannantonio}, {Goldstein},
  {Gruen}, {Gruendl}, {Guarnieri}, {Gutierrez}, {Hartley}, {Honscheid}, {Jain},
  {James}, {Jeltema}, {Jouvel}, {Kessler}, {King}, {Kirk}, {Kron}, {Kuehn},
  {Kuropatkin}, {Lahav}, {Li}, {Lima}, {Lin}, {Maia}, {Makler}, {Manera},
  {Maraston}, {Marshall}, {Martini}, {McMahon}, {Melchior}, {Merson}, {Miller},
  {Miquel}, {Mohr}, {Morice-Atkinson}, {Naidoo}, {Neilsen}, {Nichol}, {Nord},
  {Ogando}, {Ostrovski}, {Palmese}, {Papadopoulos}, {Peiris}, {Peoples},
  {Percival}, {Plazas}, {Reed}, {Refregier}, {Romer}, {Roodman}, {Ross},
  {Rozo}, {Rykoff}, {Sadeh}, {Sako}, {S{\'a}nchez}, {Sanchez}, {Santiago},
  {Scarpine}, {Schubnell}, {Sevilla-Noarbe}, {Sheldon}, {Smith}, {Smith},
  {Soares-Santos}, {Sobreira}, {Soumagnac}, {Suchyta}, {Sullivan}, {Swanson},
  {Tarle}, {Thaler}, {Thomas}, {Thomas}, {Tucker}, {Vieira}, {Vikram},
  {Walker}, {Wechsler}, {Weller}, {Wester}, {Whiteway}, {Wilcox}, {Yanny},
  {Zhang}, \& {Zuntz}}]{DES2016}
{Dark Energy Survey Collaboration}, {Abbott}, T., {Abdalla}, F.~B., {et~al.}
  2016, \mnras, 460, 1270

\bibitem[{{Davies} {et~al.}(2013){Davies}, {Agudo Berbel}, {Wiezorrek},
  {Cirasuolo}, {F{\"o}rster Schreiber}, {Jung}, {Muschielok}, {Ott}, {Ramsay},
  {Schlichter}, {Sharples}, \& {Wegner}}]{Davies2013}
{Davies}, R.~I., {Agudo Berbel}, A., {Wiezorrek}, E., {et~al.} 2013, \aap, 558,
  A56

\bibitem[{{de Jong} {et~al.}(2013){de Jong}, {Kuijken}, {Applegate}, {Begeman},
  {Belikov}, {Blake}, {Bout}, {Boxhoorn}, {Buddelmeijer}, {Buddendiek},
  {Cacciato}, {Capaccioli}, {Choi}, {Cordes}, {Covone}, {Dall'Ora}, {Edge},
  {Erben}, {Franse}, {Getman}, {Grado}, {Harnois-Deraps}, {Helmich},
  {Herbonnet}, {Heymans}, {Hildebrand t}, {Hoekstra}, {Huang}, {Irisarri},
  {Joachimi}, {K{\"o}hlinger}, {Kitching}, {La Barbera}, {Lacerda},
  {McFarland}, {Miller}, {Nakajima}, {Napolitano}, {Paolillo}, {Peacock},
  {Pila-Diez}, {Puddu}, {Radovich}, {Rifatto}, {Schneider}, {Schrabback},
  {Sifon}, {Sikkema}, {Simon}, {Sutherland}, {Tudorica}, {Valentijn}, {van der
  Burg}, {van Uitert}, {van Waerbeke}, {Veland er}, {Verdoes Kleijn}, {Viola},
  \& {Vriend}}]{deJong2013}
{de Jong}, J.~T.~A., {Kuijken}, K., {Applegate}, D., {et~al.} 2013, The
  Messenger, 154, 44

\bibitem[{{Elbaz} {et~al.}(2007){Elbaz}, {Daddi}, {Le Borgne}, {Dickinson},
  {Alexander}, {Chary}, {Starck}, {Brandt}, {Kitzbichler}, {MacDonald},
  {Nonino}, {Popesso}, {Stern}, \& {Vanzella}}]{Elbaz2007}
{Elbaz}, D., {Daddi}, E., {Le Borgne}, D., {et~al.} 2007, \aap, 468, 33

\bibitem[{{Fossati} {et~al.}(2016){Fossati}, {Fumagalli}, {Boselli}, {Gavazzi},
  {Sun}, \& {Wilman}}]{Fossati2016}
{Fossati}, M., {Fumagalli}, M., {Boselli}, A., {et~al.} 2016, \mnras, 455, 2028

\bibitem[{{Furusawa} {et~al.}(2008){Furusawa}, {Kosugi}, {Akiyama}, {Takata},
  {Sekiguchi}, {Tanaka}, {Iwata}, {Kajisawa}, {Yasuda}, {Doi}, {Ouchi},
  {Simpson}, {Shimasaku}, {Yamada}, {Furusawa}, {Morokuma}, {Ishida}, {Aoki},
  {Fuse}, {Imanishi}, {Iye}, {Karoji}, {Kobayashi}, {Kodama}, {Komiyama},
  {Maeda}, {Miyazaki}, {Mizumoto}, {Nakata}, {Noumaru}, {Ogasawara}, {Okamura},
  {Saito}, {Sasaki}, {Ueda}, \& {Yoshida}}]{Furusawa2008}
{Furusawa}, H., {Kosugi}, G., {Akiyama}, M., {et~al.} 2008, \apjs, 176, 1

\bibitem[{{Hildebrandt} {et~al.}(2010){Hildebrandt}, {Arnouts}, {Capak},
  {Moustakas}, {Wolf}, {Abdalla}, {Assef}, {Banerji}, {Ben{\'\i}tez},
  {Brammer}, {Budav{\'a}ri}, {Carliles}, {Coe}, {Dahlen}, {Feldmann}, {Gerdes},
  {Gillis}, {Ilbert}, {Kotulla}, {Lahav}, {Li}, {Miralles}, {Purger},
  {Schmidt}, \& {Singal}}]{Hildebrandt2010}
{Hildebrandt}, H., {Arnouts}, S., {Capak}, P., {et~al.} 2010, \aap, 523, A31

\bibitem[{{Ilbert} {et~al.}(2006){Ilbert}, {Arnouts}, {McCracken},
  {Bolzonella}, {Bertin}, {Le F{\`e}vre}, {Mellier}, {Zamorani}, {Pell\`o},
  {Iovino}, {Tresse}, {Le Brun}, {Bottini}, {Garilli}, {Maccagni}, {Picat},
  {Scaramella}, {Scodeggio}, {Vettolani}, {Zanichelli}, {Adami}, {Bardelli},
  {Cappi}, {Charlot}, {Ciliegi}, {Contini}, {Cucciati}, {Foucaud}, {Franzetti},
  {Gavignaud}, {Guzzo}, {Marano}, {Marinoni}, {Mazure}, {Meneux}, {Merighi},
  {Paltani}, {Pollo}, {Pozzetti}, {Radovich}, {Zucca}, {Bondi}, {Bongiorno},
  {Busarello}, {de La Torre}, {Gregorini}, {Lamareille}, {Mathez}, {Merluzzi},
  {Ripepi}, {Rizzo}, \& {Vergani}}]{Ilbert2006}
{Ilbert}, O., {Arnouts}, S., {McCracken}, H.~J., {et~al.} 2006, \aap, 457, 841

\bibitem[{{Ilbert} {et~al.}(2009){Ilbert}, {Capak}, {Salvato}, {Aussel},
  {McCracken}, {Sanders}, {Scoville}, {Kartaltepe}, {Arnouts}, {Le Floc'h},
  {Mobasher}, {Taniguchi}, {Lamareille}, {Leauthaud}, {Sasaki}, {Thompson},
  {Zamojski}, {Zamorani}, {Bardelli}, {Bolzonella}, {Bongiorno}, {Brusa},
  {Caputi}, {Carollo}, {Contini}, {Cook}, {Coppa}, {Cucciati}, {de la Torre},
  {de Ravel}, {Franzetti}, {Garilli}, {Hasinger}, {Iovino}, {Kampczyk},
  {Kneib}, {Knobel}, {Kovac}, {Le Borgne}, {Le Brun}, {Le F{\`e}vre}, {Lilly},
  {Looper}, {Maier}, {Mainieri}, {Mellier}, {Mignoli}, {Murayama}, {Pell{\`o}},
  {Peng}, {P{\'e}rez-Montero}, {Renzini}, {Ricciardelli}, {Schiminovich},
  {Scodeggio}, {Shioya}, {Silverman}, {Surace}, {Tanaka}, {Tasca}, {Tresse},
  {Vergani}, \& {Zucca}}]{Ilbert2009}
{Ilbert}, O., {Capak}, P., {Salvato}, M., {et~al.} 2009, \apj, 690, 1236

\bibitem[{{Kashino} {et~al.}(2019){Kashino}, {Silverman}, {Sanders},
  {Kartaltepe}, {Daddi}, {Renzini}, {Rodighiero}, {Puglisi}, {Valentino},
  {Juneau}, {Arimoto}, {Nagao}, {Ilbert}, {Le F{\`e}vre}, \&
  {Koekemoer}}]{Kashino2019}
{Kashino}, D., {Silverman}, J.~D., {Sanders}, D., {et~al.} 2019, \apjs, 241, 10

\bibitem[{{Kennicutt}(1998)}]{Kennicutt1998}
{Kennicutt}, Robert~C., J. 1998, \araa, 36, 189

\bibitem[{{Kitching} {et~al.}(2019){Kitching}, {Taylor}, {Capak}, {Masters}, \&
  {Hoekstra}}]{Kitching2019}
{Kitching}, T.~D., {Taylor}, P.~L., {Capak}, P., {Masters}, D., \& {Hoekstra},
  H. 2019, arXiv e-prints, arXiv:1901.06495

\bibitem[{{Kohonen}(2001)}]{Kohonen2001}
{Kohonen}, T. 2001, {Self-Organizing Maps} (Springer)

\bibitem[{{Laigle} {et~al.}(2019){Laigle}, {Davidzon}, {Ilbert}, {Devriendt},
  {Kashino}, {Pichon}, {Capak}, {Arnouts}, {de la Torre}, {Dubois},
  {Gozaliasl}, {Le Borgne}, {Lilly}, {McCracken}, {Salvato}, \&
  {Slyz}}]{Laigle2019}
{Laigle}, C., {Davidzon}, I., {Ilbert}, O., {et~al.} 2019, \mnras, 486, 5104

\bibitem[{{Laigle} {et~al.}(2016){Laigle}, {McCracken}, {Ilbert}, {Hsieh},
  {Davidzon}, {Capak}, {Hasinger}, {Silverman}, {Pichon}, {Coupon}, {Aussel},
  {Le Borgne}, {Caputi}, {Cassata}, {Chang}, {Civano}, {Dunlop}, {Fynbo},
  {Kartaltepe}, {Koekemoer}, {Le F{\`e}vre}, {Le Floc'h}, {Leauthaud}, {Lilly},
  {Lin}, {Marchesi}, {Milvang-Jensen}, {Salvato}, {Sanders}, {Scoville},
  {Smolcic}, {Stockmann}, {Taniguchi}, {Tasca}, {Toft}, {Vaccari}, \&
  {Zabl}}]{Laigle2016}
{Laigle}, C., {McCracken}, H.~J., {Ilbert}, O., {et~al.} 2016, \apjs, 224, 24

\bibitem[{{Laureijs} {et~al.}(2011){Laureijs}, {Amiaux}, {Arduini},
  {Augu{\`e}res}, {Brinchmann}, {Cole}, {Cropper}, {Dabin}, {Duvet}, {Ealet},
  {Garilli}, {Gondoin}, {Guzzo}, {Hoar}, {Hoekstra}, {Holmes}, {Kitching},
  {Maciaszek}, {Mellier}, {Pasian}, {Percival}, {Rhodes}, {Saavedra Criado},
  {Sauvage}, {Scaramella}, {Valenziano}, {Warren}, {Bender}, {Castander},
  {Cimatti}, {Le F{\`e}vre}, {Kurki-Suonio}, {Levi}, {Lilje}, {Meylan},
  {Nichol}, {Pedersen}, {Popa}, {Rebolo Lopez}, {Rix}, {Rottgering},
  {Zeilinger}, {Grupp}, {Hudelot}, {Massey}, {Meneghetti}, {Miller}, {Paltani},
  {Paulin-Henriksson}, {Pires}, {Saxton}, {Schrabback}, {Seidel}, {Walsh},
  {Aghanim}, {Amendola}, {Bartlett}, {Baccigalupi}, {Beaulieu}, {Benabed},
  {Cuby}, {Elbaz}, {Fosalba}, {Gavazzi}, {Helmi}, {Hook}, {Irwin}, {Kneib},
  {Kunz}, {Mannucci}, {Moscardini}, {Tao}, {Teyssier}, {Weller}, {Zamorani},
  {Zapatero Osorio}, {Boulade}, {Foumond}, {Di Giorgio}, {Guttridge}, {James},
  {Kemp}, {Martignac}, {Spencer}, {Walton}, {Bl{\"u}mchen}, {Bonoli},
  {Bortoletto}, {Cerna}, {Corcione}, {Fabron}, {Jahnke}, {Ligori}, {Madrid},
  {Martin}, {Morgante}, {Pamplona}, {Prieto}, {Riva}, {Toledo}, {Trifoglio},
  {Zerbi}, {Abdalla}, {Douspis}, {Grenet}, {Borgani}, {Bouwens}, {Courbin},
  {Delouis}, {Dubath}, {Fontana}, {Frailis}, {Grazian}, {Koppenh{\"o}fer},
  {Mansutti}, {Melchior}, {Mignoli}, {Mohr}, {Neissner}, {Noddle}, {Poncet},
  {Scodeggio}, {Serrano}, {Shane}, {Starck}, {Surace}, {Taylor},
  {Verdoes-Kleijn}, {Vuerli}, {Williams}, {Zacchei}, {Altieri}, {Escudero
  Sanz}, {Kohley}, {Oosterbroek}, {Astier}, {Bacon}, {Bardelli}, {Baugh},
  {Bellagamba}, {Benoist}, {Bianchi}, {Biviano}, {Branchini}, {Carbone},
  {Cardone}, {Clements}, {Colombi}, {Conselice}, {Cresci}, {Deacon}, {Dunlop},
  {Fedeli}, {Fontanot}, {Franzetti}, {Giocoli}, {Garcia-Bellido}, {Gow},
  {Heavens}, {Hewett}, {Heymans}, {Holland}, {Huang}, {Ilbert}, {Joachimi},
  {Jennins}, {Kerins}, {Kiessling}, {Kirk}, {Kotak}, {Krause}, {Lahav}, {van
  Leeuwen}, {Lesgourgues}, {Lombardi}, {Magliocchetti}, {Maguire}, {Majerotto},
  {Maoli}, {Marulli}, {Maurogordato}, {McCracken}, {McLure}, {Melchiorri},
  {Merson}, {Moresco}, {Nonino}, {Norberg}, {Peacock}, {Pello}, {Penny},
  {Pettorino}, {Di Porto}, {Pozzetti}, {Quercellini}, {Radovich}, {Rassat},
  {Roche}, {Ronayette}, {Rossetti}, {Sartoris}, {Schneider}, {Semboloni},
  {Serjeant}, {Simpson}, {Skordis}, {Smadja}, {Smartt}, {Spano}, {Spiro},
  {Sullivan}, {Tilquin}, {Trotta}, {Verde}, {Wang}, {Williger}, {Zhao},
  {Zoubian}, \& {Zucca}}]{Laureijs2011}
{Laureijs}, R., {Amiaux}, J., {Arduini}, S., {et~al.} 2011, arXiv e-prints,
  arXiv:1110.3193

\bibitem[{{Le F{\`e}vre} {et~al.}(2005){Le F{\`e}vre}, {Vettolani}, {Garilli},
  {Tresse}, {Bottini}, {Le Brun}, {Maccagni}, {Picat}, {Scaramella},
  {Scodeggio}, {Zanichelli}, {Adami}, {Arnaboldi}, {Arnouts}, {Bardelli},
  {Bolzonella}, {Cappi}, {Charlot}, {Ciliegi}, {Contini}, {Foucaud},
  {Franzetti}, {Gavignaud}, {Guzzo}, {Ilbert}, {Iovino}, {McCracken}, {Marano},
  {Marinoni}, {Mathez}, {Mazure}, {Meneux}, {Merighi}, {Paltani}, {Pell{\`o}},
  {Pollo}, {Pozzetti}, {Radovich}, {Zamorani}, {Zucca}, {Bondi}, {Bongiorno},
  {Busarello}, {Lamareille}, {Mellier}, {Merluzzi}, {Ripepi}, \&
  {Rizzo}}]{LeFevre2005}
{Le F{\`e}vre}, O., {Vettolani}, G., {Garilli}, B., {et~al.} 2005, \aap, 439,
  845

\bibitem[{{Lehmer} {et~al.}(2005){Lehmer}, {Brandt}, {Alexander}, {Bauer},
  {Schneider}, {Tozzi}, {Bergeron}, {Garmire}, {Giacconi}, {Gilli}, {Hasinger},
  {Hornschemeier}, {Koekemoer}, {Mainieri}, {Miyaji}, {Nonino}, {Rosati},
  {Silverman}, {Szokoly}, \& {Vignali}}]{Lehmer2005}
{Lehmer}, B.~D., {Brandt}, W.~N., {Alexander}, D.~M., {et~al.} 2005, \apjs,
  161, 21

\bibitem[{{Lilly} {et~al.}(2007){Lilly}, {Le F{\`e}vre}, {Renzini}, {Zamorani},
  {Scodeggio}, {Contini}, {Carollo}, {Hasinger}, {Kneib}, {Iovino}, {Le Brun},
  {Maier}, {Mainieri}, {Mignoli}, {Silverman}, {Tasca}, {Bolzonella},
  {Bongiorno}, {Bottini}, {Capak}, {Caputi}, {Cimatti}, {Cucciati}, {Daddi},
  {Feldmann}, {Franzetti}, {Garilli}, {Guzzo}, {Ilbert}, {Kampczyk}, {Kovac},
  {Lamareille}, {Leauthaud}, {Borgne}, {McCracken}, {Marinoni}, {Pello},
  {Ricciardelli}, {Scarlata}, {Vergani}, {Sanders}, {Schinnerer}, {Scoville},
  {Taniguchi}, {Arnouts}, {Aussel}, {Bardelli}, {Brusa}, {Cappi}, {Ciliegi},
  {Finoguenov}, {Foucaud}, {Franceschini}, {Halliday}, {Impey}, {Knobel},
  {Koekemoer}, {Kurk}, {Maccagni}, {Maddox}, {Marano}, {Marconi}, {Meneux},
  {Mobasher}, {Moreau}, {Peacock}, {Porciani}, {Pozzetti}, {Scaramella},
  {Schiminovich}, {Shopbell}, {Smail}, {Thompson}, {Tresse}, {Vettolani},
  {Zanichelli}, \& {Zucca}}]{Lilly2007}
{Lilly}, S.~J., {Le F{\`e}vre}, O., {Renzini}, A., {et~al.} 2007, \apjs, 172,
  70

\bibitem[{{LSST Science Collaboration} {et~al.}(2009){LSST Science
  Collaboration}, {Abell}, {Allison}, {Anderson}, {Andrew}, {Angel}, {Armus},
  {Arnett}, {Asztalos}, {Axelrod}, {Bailey}, {Ballantyne}, {Bankert},
  {Barkhouse}, {Barr}, {Barrientos}, {Barth}, {Bartlett}, {Becker}, {Becla},
  {Beers}, {Bernstein}, {Biswas}, {Blanton}, {Bloom}, {Bochanski}, {Boeshaar},
  {Borne}, {Bradac}, {Brandt}, {Bridge}, {Brown}, {Brunner}, {Bullock},
  {Burgasser}, {Burge}, {Burke}, {Cargile}, {Chand rasekharan}, {Chartas},
  {Chesley}, {Chu}, {Cinabro}, {Claire}, {Claver}, {Clowe}, {Connolly}, {Cook},
  {Cooke}, {Cooray}, {Covey}, {Culliton}, {de Jong}, {de Vries}, {Debattista},
  {Delgado}, {Dell'Antonio}, {Dhital}, {Di Stefano}, {Dickinson}, {Dilday},
  {Djorgovski}, {Dobler}, {Donalek}, {Dubois-Felsmann}, {Durech},
  {Eliasdottir}, {Eracleous}, {Eyer}, {Falco}, {Fan}, {Fassnacht}, {Ferguson},
  {Fernandez}, {Fields}, {Finkbeiner}, {Figueroa}, {Fox}, {Francke}, {Frank},
  {Frieman}, {Fromenteau}, {Furqan}, {Galaz}, {Gal-Yam}, {Garnavich},
  {Gawiser}, {Geary}, {Gee}, {Gibson}, {Gilmore}, {Grace}, {Green}, {Gressler},
  {Grillmair}, {Habib}, {Haggerty}, {Hamuy}, {Harris}, {Hawley}, {Heavens},
  {Hebb}, {Henry}, {Hileman}, {Hilton}, {Hoadley}, {Holberg}, {Holman},
  {Howell}, {Infante}, {Ivezic}, {Jacoby}, {Jain}, {R}, {Jedicke}, {Jee},
  {Garrett Jernigan}, {Jha}, {Johnston}, {Jones}, {Juric}, {Kaasalainen},
  {Styliani}, {Kafka}, {Kahn}, {Kaib}, {Kalirai}, {Kantor}, {Kasliwal},
  {Keeton}, {Kessler}, {Knezevic}, {Kowalski}, {Krabbendam}, {Krughoff},
  {Kulkarni}, {Kuhlman}, {Lacy}, {Lepine}, {Liang}, {Lien}, {Lira}, {Long},
  {Lorenz}, {Lotz}, {Lupton}, {Lutz}, {Macri}, {Mahabal}, {Mandelbaum},
  {Marshall}, {May}, {McGehee}, {Meadows}, {Meert}, {Milani}, {Miller},
  {Miller}, {Mills}, {Minniti}, {Monet}, {Mukadam}, {Nakar}, {Neill}, {Newman},
  {Nikolaev}, {Nordby}, {O'Connor}, {Oguri}, {Oliver}, {Olivier}, {Olsen},
  {Olsen}, {Olszewski}, {Oluseyi}, {Padilla}, {Parker}, {Pepper}, {Peterson},
  {Petry}, {Pinto}, {Pizagno}, {Popescu}, {Prsa}, {Radcka}, {Raddick},
  {Rasmussen}, {Rau}, {Rho}, {Rhoads}, {Richards}, {Ridgway}, {Robertson},
  {Roskar}, {Saha}, {Sarajedini}, {Scannapieco}, {Schalk}, {Schindler},
  {Schmidt}, {Schmidt}, {Schneider}, {Schumacher}, {Scranton}, {Sebag},
  {Seppala}, {Shemmer}, {Simon}, {Sivertz}, {Smith}, {Allyn Smith}, {Smith},
  {Spitz}, {Stanford}, {Stassun}, {Strader}, {Strauss}, {Stubbs}, {Sweeney},
  {Szalay}, {Szkody}, {Takada}, {Thorman}, {Trilling}, {Trimble}, {Tyson}, {Van
  Berg}, {Vand en Berk}, {VanderPlas}, {Verde}, {Vrsnak}, {Walkowicz}, {Wand
  elt}, {Wang}, {Wang}, {Warner}, {Wechsler}, {West}, {Wiecha}, {Williams},
  {Willman}, {Wittman}, {Wolff}, {Wood-Vasey}, {Wozniak}, {Young}, {Zentner},
  \& {Zhan}}]{LSST2009}
{LSST Science Collaboration}, {Abell}, P.~A., {Allison}, J., {et~al.} 2009,
  arXiv e-prints, arXiv:0912.0201

\bibitem[{{Ma} {et~al.}(2006){Ma}, {Hu}, \& {Huterer}}]{Ma2006}
{Ma}, Z., {Hu}, W., \& {Huterer}, D. 2006, \apj, 636, 21

\bibitem[{{Masters} {et~al.}(2015){Masters}, {Capak}, {Stern}, {Ilbert},
  {Salvato}, {Schmidt}, {Longo}, {Rhodes}, {Paltani}, {Mobasher}, {Hoekstra},
  {Hildebrandt}, {Coupon}, {Steinhardt}, {Speagle}, {Faisst}, {Kalinich},
  {Brodwin}, {Brescia}, \& {Cavuoti}}]{Masters2015}
{Masters}, D., {Capak}, P., {Stern}, D., {et~al.} 2015, \apj, 813, 53

\bibitem[{{Masters} {et~al.}(2017){Masters}, {Stern}, {Cohen}, {Capak},
  {Rhodes}, {Castander}, \& {Paltani}}]{Masters2017}
{Masters}, D.~C., {Stern}, D.~K., {Cohen}, J.~G., {et~al.} 2017, \apj, 841, 111

\bibitem[{{Masters} {et~al.}(2019){Masters}, {Stern}, {Cohen}, {Capak},
  {Stanford}, {Hernitschek}, {Galametz}, {Davidzon}, {Rhodes}, {Sand ers},
  {Mobasher}, {Castander}, {Pruett}, \& {Fotopoulou}}]{Masters2019}
{Masters}, D.~C., {Stern}, D.~K., {Cohen}, J.~G., {et~al.} 2019, \apj, 877, 81

\bibitem[{{Mehta} {et~al.}(2018){Mehta}, {Scarlata}, {Capak}, {Davidzon},
  {Faisst}, {Hsieh}, {Ilbert}, {Jarvis}, {Laigle}, {Phillips}, {Silverman},
  {Strauss}, {Tanaka}, {Bowler}, {Coupon}, {Foucaud}, {Hemmati}, {Masters},
  {McCracken}, {Mobasher}, {Ouchi}, {Shibuya}, \& {Wang}}]{Mehta2018}
{Mehta}, V., {Scarlata}, C., {Capak}, P., {et~al.} 2018, \apjs, 235, 36

\bibitem[{{Schlegel} {et~al.}(1998){Schlegel}, {Finkbeiner}, \&
  {Davis}}]{Schlegel1998}
{Schlegel}, D.~J., {Finkbeiner}, D.~P., \& {Davis}, M. 1998, \apj, 500, 525

\bibitem[{{Scoville} {et~al.}(2007){Scoville}, {Aussel}, {Brusa}, {Capak},
  {Carollo}, {Elvis}, {Giavalisco}, {Guzzo}, {Hasinger}, {Impey}, {Kneib},
  {LeFevre}, {Lilly}, {Mobasher}, {Renzini}, {Rich}, {Sanders}, {Schinnerer},
  {Schminovich}, {Shopbell}, {Taniguchi}, \& {Tyson}}]{Scoville2007}
{Scoville}, N., {Aussel}, H., {Brusa}, M., {et~al.} 2007, \apjs, 172, 1

\bibitem[{{Soto} {et~al.}(2017){Soto}, {de Mello}, {Rafelski}, {Gardner},
  {Teplitz}, {Koekemoer}, {Ravindranath}, {Grogin}, {Scarlata}, {Kurczynski},
  \& {Gawiser}}]{Soto2017}
{Soto}, E., {de Mello}, D.~F., {Rafelski}, M., {et~al.} 2017, \apj, 837, 6

\bibitem[{{Spergel} {et~al.}(2015){Spergel}, {Gehrels}, {Baltay}, {Bennett},
  {Breckinridge}, {Donahue}, {Dressler}, {Gaudi}, {Greene}, {Guyon}, {Hirata},
  {Kalirai}, {Kasdin}, {Macintosh}, {Moos}, {Perlmutter}, {Postman},
  {Rauscher}, {Rhodes}, {Wang}, {Weinberg}, {Benford}, {Hudson}, {Jeong},
  {Mellier}, {Traub}, {Yamada}, {Capak}, {Colbert}, {Masters}, {Penny},
  {Savransky}, {Stern}, {Zimmerman}, {Barry}, {Bartusek}, {Carpenter}, {Cheng},
  {Content}, {Dekens}, {Demers}, {Grady}, {Jackson}, {Kuan}, {Kruk}, {Melton},
  {Nemati}, {Parvin}, {Poberezhskiy}, {Peddie}, {Ruffa}, {Wallace}, {Whipple},
  {Wollack}, \& {Zhao}}]{Spergel2015}
{Spergel}, D., {Gehrels}, N., {Baltay}, C., {et~al.} 2015, arXiv e-prints,
  arXiv:1503.03757

\bibitem[{{van der Wel} {et~al.}(2014){van der Wel}, {Franx}, {van Dokkum},
  {Skelton}, {Momcheva}, {Whitaker}, {Brammer}, {Bell}, {Rix}, {Wuyts},
  {Ferguson}, {Holden}, {Barro}, {Koekemoer}, {Chang}, {McGrath},
  {H{\"a}ussler}, {Dekel}, {Behroozi}, {Fumagalli}, {Leja}, {Lundgren},
  {Maseda}, {Nelson}, {Wake}, {Patel}, {Labb{\'e}}, {Faber}, {Grogin}, \&
  {Kocevski}}]{vanderWel2014}
{van der Wel}, A., {Franx}, M., {van Dokkum}, P.~G., {et~al.} 2014, \apj, 788,
  28

\bibitem[{{Wegner} \& {Muschielok}(2008)}]{Wegner2008}
{Wegner}, M. \& {Muschielok}, B. 2008, Society of Photo-Optical Instrumentation
  Engineers (SPIE) Conference Series, Vol. 7019, {KARMA: the observation
  preparation tool for KMOS}, 70190T

\bibitem[{{Whitaker} {et~al.}(2014){Whitaker}, {Franx}, {Leja}, {van Dokkum},
  {Henry}, {Skelton}, {Fumagalli}, {Momcheva}, {Brammer}, {Labb{\'e}},
  {Nelson}, \& {Rigby}}]{Whitaker2014}
{Whitaker}, K.~E., {Franx}, M., {Leja}, J., {et~al.} 2014, \apj, 795, 104

\bibitem[{{Wilkinson} {et~al.}(2015){Wilkinson}, {Maraston}, {Thomas},
  {Coccato}, {Tojeiro}, {Cappellari}, {Belfiore}, {Bershady}, {Blanton},
  {Bundy}, {Cales}, {Cherinka}, {Drory}, {Emsellem}, {Fu}, {Law}, {Li},
  {Maiolino}, {Masters}, {Tremonti}, {Wake}, {Wang}, {Weijmans}, {Xiao}, {Yan},
  {Zhang}, {Bizyaev}, {Brinkmann}, {Kinemuchi}, {Malanushenko}, {Malanushenko},
  {Oravetz}, {Pan}, \& {Simmons}}]{Wilkinson2015}
{Wilkinson}, D.~M., {Maraston}, C., {Thomas}, D., {et~al.} 2015, \mnras, 449,
  328

\bibitem[{{Wilman} {et~al.}(2020){Wilman}, {Fossati}, {Mendel}, {Saglia},
  {Wisnioski}, {Wuyts}, {Schreiber}, {Beifiori}, {Bender}, {Belli},
  {{\"U}bler}, {Lang}, {Chan}, {Davies}, {Nelson}, {Genzel}, {Tacconi},
  {Galametz}, {Davies}, {Lutz}, {Price}, {Burkert}, {Tadaki}, {Herrera-Camus},
  {Brammer}, {Momcheva}, \& {Dokkum}}]{Wilman2020}
{Wilman}, D.~J., {Fossati}, M., {Mendel}, J.~T., {et~al.} 2020, \apj, 892, 1

\bibitem[{{Wisnioski} {et~al.}(2019){Wisnioski}, {F{\"o}rster Schreiber},
  {Fossati}, {Mendel}, {Wilman}, {Genzel}, {Bender}, {Wuyts}, {Davies},
  {{\"U}bler}, {Bandara}, {Beifiori}, {Belli}, {Brammer}, {Chan}, {Davies},
  {Fabricius}, {Galametz}, {Lang}, {Lutz}, {Nelson}, {Momcheva}, {Price},
  {Rosario}, {Saglia}, {Seitz}, {Shimizu}, {Tacconi}, {Tadaki}, {van Dokkum},
  \& {Wuyts}}]{Wisnioski2019}
{Wisnioski}, E., {F{\"o}rster Schreiber}, N.~M., {Fossati}, M., {et~al.} 2019,
  \apj, 886, 124

\bibitem[{{Wright} {et~al.}(2019){Wright}, {Hildebrandt}, {van den Busch}, \&
  {Heymans}}]{Wright2019}
{Wright}, A.~H., {Hildebrandt}, H., {van den Busch}, J.~L., \& {Heymans}, C.
  2019, arXiv e-prints, arXiv:1909.09632

\bibitem[{{Wuyts} {et~al.}(2013){Wuyts}, {F{\"o}rster Schreiber}, {Nelson},
  {van Dokkum}, {Brammer}, {Chang}, {Faber}, {Ferguson}, {Franx}, {Fumagalli},
  {Genzel}, {Grogin}, {Kocevski}, {Koekemoer}, {Lundgren}, {Lutz}, {McGrath},
  {Momcheva}, {Rosario}, {Skelton}, {Tacconi}, {van der Wel}, \&
  {Whitaker}}]{Wuyts2013}
{Wuyts}, S., {F{\"o}rster Schreiber}, N.~M., {Nelson}, E.~J., {et~al.} 2013,
  \apj, 779, 135

\end{thebibliography}

\end{document}